\documentclass[iop,apj,tighten]{emulateapj}

\usepackage{apjfonts}
\usepackage{graphicx}
\usepackage{dcolumn}
\usepackage{bm}
\usepackage{epsfig}

\usepackage{float}
\usepackage{amsmath,amstext}
\usepackage{floatflt}
\usepackage{subfigure}
\usepackage[usenames]{color}
\usepackage{ulem}
\usepackage{ragged2e}

\usepackage[breaklinks,colorlinks,citecolor=blue,linkcolor=magenta]{hyperref}
\usepackage[all]{hypcap} 

\shorttitle{Tilted flat and untilted nonflat XCDM parameterizations}
\shortauthors{Park \& Ratra}

\begin{document}


\title{Observational constraints on the tilted flat-XCDM and the untilted nonflat XCDM dynamical dark energy inflation parameterizations}

\author{
Chan-Gyung Park\altaffilmark{1, 2} and
Bharat Ratra\altaffilmark{2}
}

\altaffiltext{1}{Division of Science Education and Institute of Fusion
                 Science, Chonbuk National University, Jeonju 54896, South Korea;
                 e-mail: park.chan.gyung@gmail.com}
\altaffiltext{2}{Department of Physics, Kansas State University, 116 Cardwell Hall,
                 Manhattan, KS 66506, USA
                 }

\date{\today}


\keywords{cosmological parameters --- cosmic background radiation --- large-scale structure of universe --- inflation --- observations --- methods:statistical}

%
%
\begin{abstract}
We constrain tilted spatially-flat and untilted nonflat XCDM dynamical 
dark energy inflation parameterizations using Planck 2015 cosmic 
microwave background (CMB) anisotropy data and recent baryonic 
acoustic oscillations distance measurements, Type Ia supernovae 
data, Hubble parameter observations, and growth rate measurements. 
Inclusion of the four non-CMB data sets results in a significant 
strengthening of the evidence for nonflatness in the nonflat XCDM 
model from 1.1$\sigma$ for the CMB data alone to 3.4$\sigma$ for the 
full data combination. In this untilted nonflat XCDM case the data 
favor a spatially-closed model in which spatial curvature contributes 
a little less than a percent of the current cosmological energy 
budget; they also mildly favor dynamical dark energy over a 
cosmological constant at 1.2$\sigma$. These data are also better fit 
by the flat-XCDM parameterization than by the standard $\Lambda$CDM
model, but only at 0.3$\sigma$ significance. Current data is unable 
to rule out dark energy dynamics. The nonflat XCDM parameterization 
is compatible with the Dark Energy Survey limits 
on the present value of the rms mass fluctuations amplitude 
($\sigma_8$) as a function of the present value of the nonrelativistic 
matter density parameter ($\Omega_m$), however it does not provide as good a 
fit to the higher multipole CMB temperature anisotropy data as does 
the standard tilted flat-$\Lambda$CDM model. A number of measured 
cosmological parameter values differ significantly when determined 
using the tilted flat-XCDM and the nonflat XCDM parameterizations, 
including the baryonic matter density parameter and the reionization 
optical depth.       
\end{abstract}

\maketitle

%
%

\section{Introduction}

In the standard spatially-flat $\Lambda$CDM cosmological model
\citep{Peebles1984} the current cosmological energy budget is dominated 
by the cosmological constant $\Lambda$ which powers the currently 
accelerating cosmological expansion. Cold dark matter (CDM) and baryonic 
matter are the next two largest contributors to the current 
energy budget, followed by small 
contributions from neutrinos and photons. For reviews of the standard model 
see \citet{RatraVogeley2008}, \citet{Martin2012}, \citet{Brax2018}, and
\citet{Lukovicetal2018}. This model is able to accommodate most 
observational constraints, including CMB anisotropy measurements 
\citep{PlanckCollaboration2016}, baryonic acoustic oscillation (BAO) 
distance observations \citep{Alametal2017}, Hubble parameter data 
\citep{Farooqetal2017},\footnote{Hubble parameter values have been 
measured from low redshift to well past the redshift of the 
cosmological deceleration-acceleration transition between the earlier 
nonrelativistic-matter-dominated decelerating 
cosmological expansion and the more recent dark-energy-dominated 
accelerating cosmological expansion. The transition redshift has 
been measured from Hubble parameter observations and it is 
roughly at the value expected in dark energy models \citep{FarooqRatra2013, Morescoetal2016, Farooqetal2017, Yuetal2018}.} 
and Type Ia supernova (SNIa) apparent magnitude measurements 
\citep{Scolnicetal2017}. Current observational constraints however allow
for slightly nonflat spatial geometries and/or mild dark energy 
dynamics.

The standard spatially-flat $\Lambda$CDM inflation model is 
characterized by six cosmological parameters conventionally chosen to be:
$\Omega_{b} h^2$ and $\Omega_{c} h^2$, the current values of the baryonic
and cold dark matter density parameters multiplied by the square of the 
Hubble constant $H_0$ (in units of 100 km s$^{-1}$  Mpc$^{-1}$);
$\tau$, the reionization optical depth;
$\theta_{\rm MC}$, the angular diameter distance as a multiple of the sound 
horizon at recombination; and $n_{s}$ and $A_{s}$, the 
spectral index and amplitude of the power-law primordial scalar 
fractional energy density spatial inhomogeneity power spectrum.

Observational data are on the verge of being able to place interesting 
constraints on seven parameter cosmological models. Two more plausible seventh
cosmological parameters now under discussion are spatial curvature in nonflat 
extensions of the standard model and a parameter that governs dark energy 
dynamics in dynamical dark energy extensions of the standard model. 

A simple, and so widely used, dynamical dark energy parameterization is 
the XCDM one.\footnote{Many observations have been used to 
constrain the XCDM parameterization \citep[see, e.g.,][and references therein]{ChenRatra2004, Samushiaetal2007, SamushiaRatra2010, ChenRatra2011b, Solaetal2017a, Solaetal2018, Solaetal2017b, Solaetal2017c, Solaetal2017d, Zhaietal2017}.}  
This parameterizes the equation of state relation between 
the pressure and energy density of the dark energy fluid through 
$p_X = w \rho_X$ where the equation of state parameter $w$ is the 
additional seventh cosmological parameter. XCDM is not a physically consistent 
description of dark energy as it is unable to consistently describe 
the evolution of energy density spatial inhomogeneities. To render XCDM 
physically consistent requires an eighth cosmological parameter, the 
square of the speed of sound in the dark energy fluid, $c_{sX}^2 = 
dp_X/d\rho_X$. In this paper, as is common practice, we consider a 
restricted, physically-consistent, modified XCDM parameterization in which 
$c_{sX}^2$ is not allowed to vary in space or with time and is arbitrarily 
set to unity.  
$\phi$CDM is the simplest physically consistent dynamical dark energy model 
\citep{PeeblesRatra1988, RatraPeebles1988}. Here a scalar field $\phi$ with 
potential energy density $V(\phi) \propto \phi^{-\alpha}$ is the dynamical 
dark energy with $\alpha > 0$ being the additional seventh cosmological 
parameter.\footnote{While XCDM is widely used to model dynamical dark energy, 
it does not accurately model $\phi$CDM \citep{PodariuRatra2001, Oobaetal2018d}.} 

\citet{Oobaetal2018d} \citep[also see][]{ParkRatra2018b} have analyzed the Planck 2015 CMB anisotropy data and 
some BAO distance measurements by using these seven parameter tilted 
spatially-flat XCDM and $\phi$CDM dynamical dark energy inflation models and
found that both were slightly favored by the data, compared to the standard
six parameter flat-$\Lambda$CDM model, by 1.1$\sigma$ (1.3$\sigma$) for 
the XCDM ($\phi$CDM) case. While these are not significant improvements
over the standard model, current data are not able to rule out dark energy 
dynamics. In 
addition, both dynamical dark energy models reduce the tension between 
the Planck 2015 CMB anisotropy and the weak gravitational lensing 
constraints on $\sigma_8$, the rms fractional energy density spatial inhomogeneity 
averaged over 8$h^{-1}$ Mpc radius spheres, 

There have been a number of earlier suggestions that different combinations 
of observational data favor dynamical dark energy models over the standard 
$\Lambda$CDM model \citep{Sahnietal2014, Dingetal2015, Solaetal2015, Zhengetal2016, Solaetal2017a, Solaetal2018, Solaetal2017b, Zhaoetal2017, Solaetal2017c, Zhangetal2017a, Solaetal2017d, GomezValentSola2017, Caoetal2018, GomezValentSola2018}. As far as we are aware, of these analyses, only those of \citet{Zhaoetal2017} and \citet{Zhangetal2017a} performed complete CMB anisotropy analyses of the 
generalized XCDM dynamical dark energy parameterizations they 
assumed.\footnote{Both analyses also included in their data compilation a 
high value of $H_0$ estimated from the local expansion rate. We do not include 
this high local $H_0$ value in the data compilation we use here to constrain 
cosmological parameters, as it is not consistent with the other data we use, 
in the $\Lambda$CDM, XCDM, and $\phi$CDM models.}
The other analyses either ignored CMB anisotropy data or only approximately
accounted for it.

The standard $\Lambda$CDM model assumes flat spatial hypersurfaces. In nonflat
models non-vanishing spatial curvature introduces a new length 
scale and so it is incorrect to assume a power spectrum for 
energy density inhomogeneities in nonflat models that does not 
correctly account for the spatial curvature length scale \citep[as was assumed 
for analyses of nonflat models in][]{PlanckCollaboration2016}. Nonflat
cosmological inflation provides the only known method for computing a 
physically consistent power spectrum in nonflat models. For open 
spatial hypersurfaces the \cite{Gott1982} open-bubble inflation model 
is used to compute the non-power-law power spectrum 
\citep{RatraPeebles1994, RatraPeebles1995}. For closed spatial hypersurfaces 
Hawking's prescription for the initial quantum state of the universe 
\citep{Hawking1984, Ratra1985} is used to define a closed 
inflation model that gives the non-power-law power spectrum
of spatial inhomogeneities \citep{Ratra2017}. In the nonflat inflation models,
compared to the flat inflation model, there is no simple way to allow for 
tilt so $n_s$ is not a free parameter and it is replaced by the present value 
of the spatial curvature density parameter $\Omega_k$.

Using such a physically consistent untilted nonflat inflation model 
non-power-law power 
spectrum of energy density inhomogeneities, \citet{Oobaetal2018a} found that 
Planck 2015 CMB data \citep{PlanckCollaboration2016} do not 
require flat spatial hypersurfaces in the six parameter nonflat $\Lambda$CDM 
model.\footnote{Non-CMB observations do not tightly constrain spatial 
curvature \citep{Farooqetal2015, Caietal2016, Chenetal2016, YuWang2016, LHuillierShafieloo2017, Farooqetal2017, Lietal2016, WeiWu2017, Ranaetal2017, Yuetal2018, Mitraetal2018, Mitraetal2019, Ryanetal2018, Ryanetal2019}, except for a compilation of all of the most 
recent SNIa, BAO and Hubble parameter data, which also (mildly) favors closed 
spatial hypersurfaces \citep{ParkRatra2018a} and for a compilation of 
primordial deuterium abundance measurements which favors a flat geometry
\citep{Pentonetal2018}.}
In the six parameter nonflat $\Lambda$CDM model, compared to the 
six parameter flat-$\Lambda$CDM model, $n_s$ is replaced by $\Omega_k$.
\citet{ParkRatra2018a} used the largest compilation of current reliable 
observational data to study the nonflat $\Lambda$CDM inflation model,
confirming the results of \citet{Oobaetal2018a} and finding stronger 
evidence for nonflatness, 5.1$\sigma$, favoring a very slightly 
closed model. The CMB anisotropy measurements also do not demand flat spatial 
hypersurfaces in the 
seven parameter nonflat XCDM dynamical dark energy inflation parameterization 
\citep{Oobaetal2018b}. Here $w$ is the seventh cosmological parameter and
again $n_s$ is replaced by $\Omega_k$. In the simplest seven 
parameter nonflat $\phi$CDM dynamical dark energy inflation model 
\citep{Pavlovetal2013} --- in which $\alpha$ is the seventh cosmological
parameter with $n_s$ replaced by $\Omega_k$ --- \citet{Oobaetal2018c} 
\citep[also see][]{ParkRatra2018b} again found that CMB anisotropy 
observations do not require flat spatial geometry.
In both the XCDM and $\phi$CDM inflation cases the data also favor a very 
slightly closed model. All three slightly closed models are more consistent 
with $\sigma_8$ constraints from weak lensing observations.

In this paper we determine observational constraints on the seven parameter
tilted flat-XCDM\footnote{For earlier constraints on the flat-XCDM model, see \cite{Zhaoetal2007}, \cite{Wangetal2007}, \cite{Wangetal2009}, and references therein.} and the seven parameter untilted nonflat XCDM dynamical 
dark energy 
inflation parameterizations. For this purpose we use an updated version of the 
Planck 2015 CMB anisotropy, and (almost all currently available reliable) 
SNIa apparent magnitude, BAO distance, growth factor, and Hubble parameter 
data compilation of \cite{ParkRatra2018a}. Our main update here is the
replacement of the Joint Light-curve Analysis (JLA) compilation of 
apparent magnitude measurements of 740 SNIa \citep{Betouleetal2014} by the 
Pantheon collection of 1048 SNIa measurements \citep{Scolnicetal2017}.
When used with the Planck 2015 CMB anisotropy data
in an analysis of the nonflat $\Lambda$CDM case, the Pantheon data place
tighter constraints on spatial curvature than do the JLA data. Overall,
for the full data compilation, our updated results for the tilted 
flat-$\Lambda$CDM inflation model and the nonflat $\Lambda$CDM inflation 
model here are 
very similar to those of \citet{ParkRatra2018a}, with evidence for 
nonflatness in the nonflat $\Lambda$CDM case now becoming 5.2$\sigma$ 
(from 5.1$\sigma$).

Our first main goal here is to examine the consequences of including a 
significant amount of recent, reliable, non-CMB data on the discovery of 
\citet{Oobaetal2018d} that the Planck 2015 CMB anisotropy data and 
a few BAO distance measurements favor the seven parameter 
tilted flat-XCDM parameterization over the six parameter standard 
flat-$\Lambda$CDM model.
Our second main goal is to examine the effect that the inclusion of this 
new non-CMB data compilation has on the discovery of \citet{Oobaetal2018b} 
that the Planck 2015 CMB anisotropy observations and a few BAO distance 
measurements are
not inconsistent with the closed-XCDM inflation parameterization. Our 
third main goal is to use this large compilation of reliable data 
to examine the compatibility of the cosmological constraints from each 
type of data and to also more tightly measure cosmological parameters than 
has been achieved to date, and in particular to also find out which
model parameters can or cannot be measured from these data in a 
cosmological-model-independent manner.   

We find that the seven parameter tilted flat-XCDM inflation parameterization
continues to provide a better fit to the data than does the six parameter
standard $\Lambda$CDM model. However, for the large compilation of data used
here we find the XCDM dynamical dark energy case is only 0.28$\sigma$ better 
than the standard $\Lambda$CDM case (compared to the 1.1$\sigma$ 
\citealt{Oobaetal2018d} found with the smaller data compilation). This is
not a significant improvement over the standard model but on the other hand 
the XCDM parameterization cannot be ruled out.
Also in agreement with \citet{Oobaetal2018d} we do not find a deviation
from $w = -1$ (a cosmological constant) for the flat-XCDM case.\footnote{These
results differ from those of earlier approximate analyses, based on less
and less reliable data than we have used here \citep{Solaetal2017a, Solaetal2018,
Solaetal2017b, Solaetal2017c, Solaetal2017d, GomezValentSola2017, GomezValentSola2018},
that favor the flat-XCDM case over the flat-$\Lambda$CDM one by 3$\sigma$ or
greater and find $w$ deviating from $-1$ by more than 3$\sigma$.}
Similar to the $\Lambda$CDM models in \citet{ParkRatra2018a},
the tilted flat-XCDM model continues to better fit the weak lensing bounds
in the $\sigma_8$--$\Omega_m$ plane than does the untilted nonflat XCDM model.

For the untilted nonflat XCDM inflation parameterization, our results here, determined
using much more non-CMB data, are consistent with and strengthen those of 
\citet{Oobaetal2018b}. For the full data compilation we now find a more than 
3.4$\sigma$ deviation from spatial flatness and now, for the first time, 
we also find a corresponding deviation from a cosmological constant with
$w = -0.960 \pm 0.032$ in the nonflat XCDM case being more than 1.2$\sigma$ 
away from the cosmological constant value of $w = -1$. The nonflat XCDM 
parameterization better fits the weak lensing limits in the 
$\sigma_8$--$\Omega_m$ plane. For the full data combination we consider here
(including CMB lensing data), we find that the observed low-$\ell$ CMB 
temperature and polarization anisotropy multipole number ($\ell$) power 
spectrum $C_\ell$ is best 
fit\footnote{Here by best fit we mean that the corresponding model has the 
lowest $\chi^2$ of the models under consideration. As discussed elsewhere 
and below, a number of these models are 
not nested and the Planck 2015 data number of degrees of freedom are 
ambiguous, so in many cases it is not possible to convert the 
$\Delta \chi^2$'s we compute here to a quantitative goodness of fit and so 
many of our statements here about goodness of fit are qualitative. See below 
for more detailed discussion of this issue.}
by the tilted flat-$\Lambda$CDM model, followed by the tilted 
flat-XCDM parameterization, with the untilted nonflat $\Lambda$CDM 
model and the 
untilted nonflat XCDM parameterization in third and fourth place. The 
tilted flat-$\Lambda$CDM model and the tilted flat-XCDM parameterization
best fit the observed higher-$\ell$ $C_\ell$'s, followed by the untilted 
nonflat XCDM parameterization and the untilted nonflat $\Lambda$CDM model
in third and fourth place.    

We find that $H_0$ is measured in an almost model-independent 
manner with values that are consistent with most other estimates. However, 
as also found in \citet{ParkRatra2018a}, some
measured cosmological parameter values, including $\Omega_{b} h^2$,
$\tau$, and $\Omega_{c} h^2$, differ significantly between the flat and 
the nonflat models and so caution is needed when utilizing cosmological 
measurements of such parameters. 

In Sec.\ 2 we briefly summarize the 
cosmological data we use in our analyses. In Sec.\ 3 we briefly 
summarize the methods we use for our analyses. The observational 
constraints resulting 
from these data for the tilted flat-XCDM and the nonflat 
XCDM inflation parameterizations are presented in Sec.\ 4. We conclude 
in Sec.\ 5.  

\section{Data}

As in \citet{ParkRatra2018a} we use the TT + lowP and 
TT + lowP + lensing Planck 2015 CMB anisotropy data \citep{PlanckCollaboration2016}. 
Here TT denotes the low-$\ell$
($2 \le \ell \le 29$) and high-$\ell$ ($30 \le \ell \le 2508$; PlikTT)
Planck 2015 temperature-only $C_\ell^{TT}$ data and lowP represents low-$\ell$
polarization $C_\ell^{TE}$, $C_\ell^{EE}$, and $C_\ell^{BB}$ power
spectra measurements at $2 \le \ell \le 29$. The collection of low-$\ell$
temperature and polarization measurements is referred to as lowTEB.
The CMB lensing data we use is the power spectrum of the lensing potential
measured by Planck.

Instead of using the JLA apparent magnitude measurement compilation of 740 SNIa 
\citep{Betouleetal2014}, we use the Pantheon collection of 1048 SNIa apparent 
magnitude measurements over the broader redshift range of $0.01 < z < 2.3$ 
\citep{Scolnicetal2017}, which includes 276 SNIa ($0.03 < z < 0.65$)
discovered by the Pan-STARRS1 Medium Deep Survey and SNIa distance estimates
from SDSS, SNLS and low-$z$ HST samples. Throughout this paper, we use the 
abbreviation SN to denote the Pantheon SNIa sample. 

We make one change to the BAO compilation of Sec.\ 2.3 and Table 1 of 
\citet{ParkRatra2018a}, here using 
$D_V (r_{d,\textrm{fid}} / r_d)=3843\pm147$ Mpc for the \cite{Ataetal2018} BAO 
data point, instead of the old value, 
$D_V (r_{d,\textrm{fid}} / r_d)=3855\pm170$ Mpc,
given in the initial version of their preprint (arXiv:1705.06373v1). 
See Sec.\ 2.3 of \citet{ParkRatra2018a} for the definitions of the above 
expressions. Throughout this paper, we use the abbreviation BAO 
to denote this updated BAO compilation.

We also use the same Hubble parameter, $H(z)$, and growth rate, 
$f(z) \sigma_8 (z)$, measurements listed in Tables 2 and 3 of 
\citet{ParkRatra2018a}. More precisely, Table 2 of 
\citet{ParkRatra2018a} lists all more reliable cosmic chronometric 
$H(z)$ data (31 measurements in  all), while Table 1 of 
\citet{ParkRatra2018a} includes three radial BAO $H(z)$ meaurements.

\section{Methods}
\label{sec:methods}

We use the publicly available CAMB/COSMOMC analysis software (November 2016 
version) \citep{ChallinorLasenby1999, Lewisetal2000, LewisBridle2002}
to constrain cosmological parameters of the tilted flat and the untilted 
nonflat XCDM dynamical dark energy inflation parameterizations with Planck 
2015 CMB measurements and non-CMB data sets. We use the CAMB Boltzmann 
code to compute the angular power spectra for CMB temperature 
fluctuations, polarization, and lensing potential, and COSMOMC, based
on the Markov chain Monte Carlo (MCMC) method, to determine the range of 
model parameters favored by the data. We use the same COSMOMC settings 
adopted in \citet{PlanckCollaboration2016} and used in \citet{ParkRatra2018a}. 

The spatially-flat tilted XCDM inflation case primordial power spectrum 
\citep{LucchinMatarrese1985, Ratra1992, Ratra1989} is
\begin{equation}
   P(k)=A_s \left(\frac{k}{k_0} \right)^{n_s},
\end{equation}
where $A_s$ is the amplitude of the power spectrum at the pivot scale 
$k_0=0.05~\textrm{Mpc}^{-1}$ 
and $k$ is wavenumber. The untilted nonflat XCDM inflation case 
primordial power spectrum \citep{RatraPeebles1995, Ratra2017} is
\begin{equation}
   P(q) \propto \frac{(q^2-4K)^2}{q(q^2-K)},
\end{equation}
which reduces to the $n_s=1$ spectrum in the spatially-flat limit ($K=0$).
For scalar perturbations, $q=\sqrt{k^2 + K}$ is wavenumber where 
$K=-(H_0^2 / c^2 ) \Omega_k$ is spatial curvature and $c$ is the 
speed of light. In the spatially-closed model, with negative $\Omega_k$, 
normal modes are characterized by positive integers 
$\nu=q K^{-1/2}=3,4,5,\cdots$. We use $P(q)$ as the initial spatial 
inhomogeneity perturbation power spectrum for the nonflat model by
normalizing it at the pivot scale $k_0$ to the value of $A_s$. 

Our analyses methods are very similar to those described in Sec.\ 3.2
of \citet{ParkRatra2018a}. During the MCMC process we set the same priors 
for the cosmological parameters as in \citet{ParkRatra2018a}. For the seventh 
parameter $w$, we set $-3\le w\le 0.2$.
In our analyses with the Pantheon SNIa sample, we do not set priors
for the nuisance parameters ($\alpha_\textrm{SN}$ and $\beta_\textrm{SN}$)
related to the stretch and the color correction of the SNIa light curves,
since the stretch and color parameters of the Pantheon SNIa used here are
set to zero.\footnote{In addition to $\alpha_\textrm{SN}$ and $\beta_\textrm{SN}$,
the distance moduli of the Pantheon SNIa are affected by three more nuisance parameters,
the absolute $B$-band magnitude ($M_B$), the distance correction based on the host-galaxy
mass ($\Delta_M$), and the distance correction based on predicted biases from 
simulation ($\Delta_B$) \citep{Scolnicetal2017}. Consequently, the number
of degrees of freedom of the Pantheon sample is less than the number of SNIa.
For example, for a flat-$\Lambda\textrm{CDM}$ model analysis that fits
$\Omega_m$, the number of degrees
of freedom becomes 1042 ($=1048-6$).}

\section{Observational Constraints}

\begin{figure*}
\centering
\mbox{\includegraphics[width=87mm]{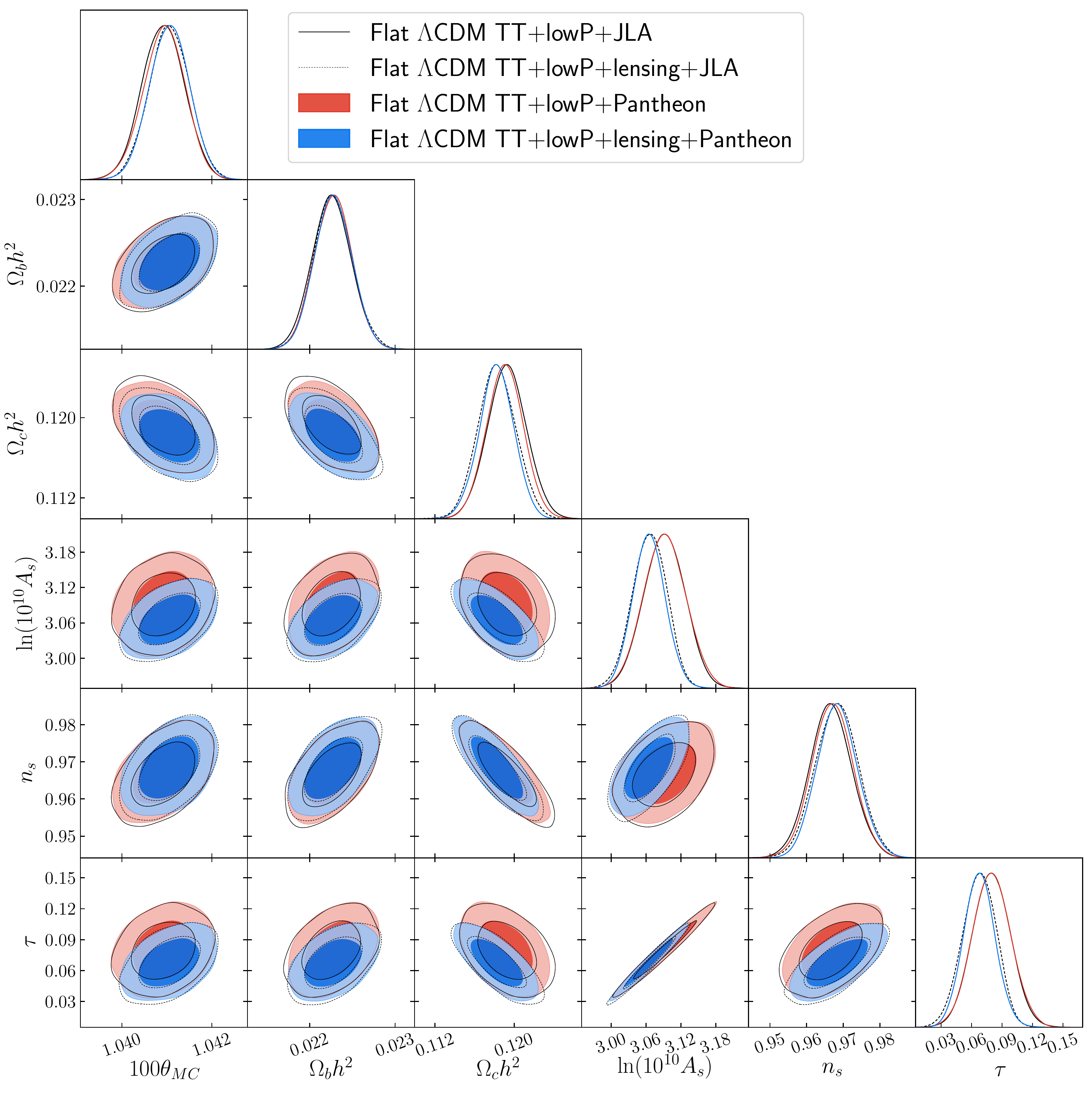}}
\mbox{\includegraphics[width=87mm]{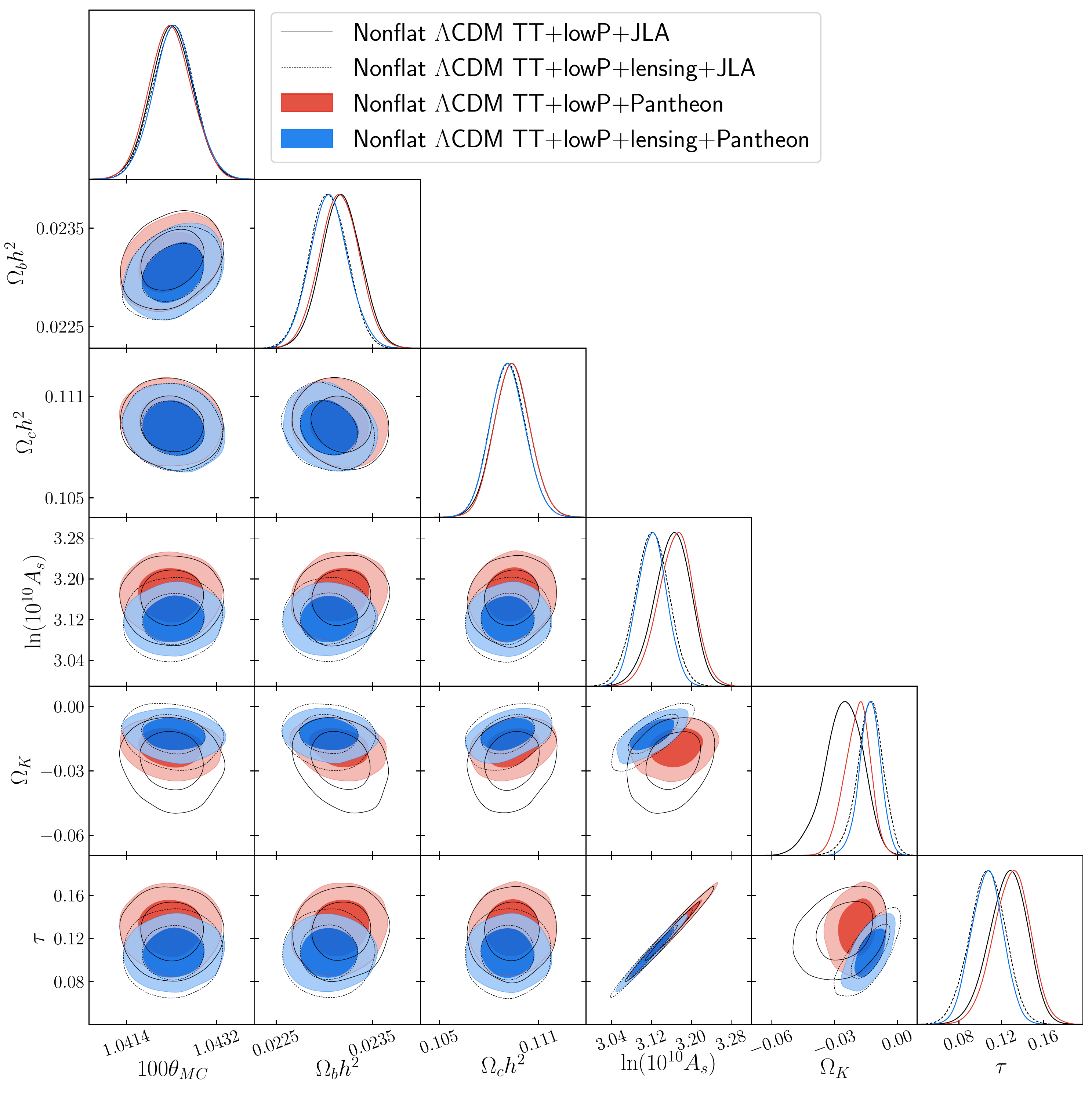}}
\caption{Likelihood distributions of the tilted flat (left) and untilted 
nonflat (right) $\Lambda\textrm{CDM}$ inflation model parameters favored 
by the Planck
CMB TT + lowP (+ lensing) and SNIa data. Here the parameter constraints are
compared for the JLA SNIa data and the Pantheon SNIa data and summarized
in Table \ref{tab:para_SN}. Two-dimensional marginalized likelihood 
contours as well as one-dimensional likelihoods
are shown as solid and dashed black curves for JLA and filled contours and colored
curves for Pantheon data.
}
\label{fig:jla_pantheon}
\end{figure*}

We first examine how much more effective the improved Pantheon SNIa data are
in constraining cosmological parameters, relative to the JLA data. Figure 
\ref{fig:jla_pantheon} compares the likelihood distributions of the model 
parameters for the JLA and the Pantheon data sets, in conjunction with the 
CMB observations, for the spatially-flat tilted and for the untilted nonflat 
$\Lambda$CDM inflation models. The mean and 68.3\% confidence limits of 
model parameters are presented in Table \ref{tab:para_SN}.\footnote{The 
parameter values of the tilted flat-$\Lambda\textrm{CDM}$ model constrained 
by using TT + lowP (+lensing) + JLA data are in good agreement with the 
Planck results. See Planck 2015 cosmological parameter tables 
base\_plikHM\_TT\_lowTEB\_post\_JLA
for TT + lowP + JLA data and base\_plikHM\_TT\_lowTEB\_lensing\_post\_JLA
for TT + lowP + lensing + JLA data \citep{PlanckParameterTables2015}.} 
Without CMB lensing data, the Pantheon data are a little more constraining
than the JLA data. When CMB lensing data are included, the largest reduction
in error bars occur for the nonflat
$\Lambda$CDM case, where the error bars for $\Omega_m$, $H_0$, and $\Omega_k$
are approximately only 80\% as large for the CMB + Pantheon combination 
when compared to the CMB + JLA case. From this Table we also see that 
including CMB lensing measurements results in a decrease of $A_s$ 
and $\tau$ in both models.

Note that our six parameter physically-consistent untilted non-power-law power
spectrum nonflat $\Lambda\textrm{CDM}$ model constrained by the CMB and 
Pantheon data favors nonflat geometry and that the parameter constraints 
determined using our model are quite different from those presented in 
\citet{Scolnicetal2017} that were derived using the seven parameter 
physically-inconsistent tilted power-law power spectrum nonflat $\Lambda$CDM
model
(with varying spectral index $n_s$) which were found to favor spatial flatness
($\Omega_k=0.004\pm 0.006$, $\Omega_m=0.295\pm0.024$, $H_0=69.695\pm2.933~\textrm{km}~\textrm{s}^{-1}~\textrm{Mpc}^{-1}$ for the TT + lowP + Pantheon data 
combination).

\begin{table*}
\caption{Mean and 68.3\% confidence limits of tilted flat and untilted nonflat $\Lambda\textrm{CDM}$ model parameters constrained by Planck and SNIa data. JLA versus Pantheon.}
\begin{ruledtabular}
\begin{tabular}{lcccc}
 \multicolumn{5}{c}{Tilted flat-$\Lambda\textrm{CDM}$ model} \\
 \hline \\[-2mm]
  Parameter               & TT+lowP+JLA            &  TT+lowP+lensing+JLA    &  TT+lowP+Pantheon       &  TT+lowP+lensing+Pantheon \\[+0mm]
 \hline \\[-2mm]
  $\Omega_b h^2$          & $0.02226 \pm 0.00023$  &  $0.02227 \pm 0.00022$  &  $0.02228 \pm 0.00022$  &  $0.02228 \pm 0.00022$   \\[+1mm]
  $\Omega_c h^2$          & $0.1193  \pm 0.0020$   &  $0.1183  \pm 0.0019$   &  $0.1191  \pm 0.0019$   &  $0.1182  \pm 0.0017$    \\[+1mm]
  $100\theta_\textrm{MC}$ & $1.04092 \pm 0.00047$  &  $1.04105 \pm 0.00045$  &  $1.04094 \pm 0.00046$  &  $1.04106 \pm 0.00044$   \\[+1mm]
  $\tau$                  & $0.080   \pm 0.019$    &  $0.068   \pm 0.016$    &  $0.080   \pm 0.019$    &  $0.068   \pm 0.015$     \\[+1mm]
  $\ln(10^{10} A_s)$      & $3.092   \pm 0.035$    &  $3.066   \pm 0.029$    &  $3.092   \pm 0.036$    &  $3.065   \pm 0.028$     \\[+1mm]
  $n_s$                   & $0.9666  \pm 0.0057$   &  $0.9683  \pm 0.0058$   &  $0.9671  \pm 0.0056$   &  $0.9684  \pm 0.0055$    \\[+1mm]
 \hline \\[-2mm]
  $H_0$ [km s$^{-1}$ Mpc$^{-1}$] & $67.52 \pm0.89$ &  $67.93 \pm 0.88$       &  $67.62   \pm 0.84$     &  $67.96   \pm 0.80$      \\[+1mm]
  $\Omega_m$              & $0.312 \pm0.012$       &  $0.306  \pm 0.012$     &  $0.311  \pm 0.011$     &  $0.306   \pm 0.011$    \\[+1mm]
  $\sigma_8$              & $0.829  \pm 0.014$     &  $0.8156  \pm 0.0093$   &  $0.829  \pm 0.015$     &  $0.8152  \pm 0.0094$    \\[+1mm]
    \hline \hline \\[-2mm]
   \multicolumn{5}{c}{Untilted nonflat $\Lambda\textrm{CDM}$ model} \\
   \hline \\[-2mm]
  Parameter               & TT+lowP+JLA            &  TT+lowP+lensing+JLA     &  TT+lowP+Pantheon      &  TT+lowP+lensing+Pantheon   \\[+0mm]
 \hline \\[-2mm]
  $\Omega_b h^2$          & $0.02318 \pm0.00020$   &  $0.02304 \pm 0.00020$  &  $0.02316 \pm 0.00020$  &  $0.02305 \pm 0.00020$   \\[+1mm]
  $\Omega_c h^2$          & $0.1094  \pm0.0011$    &  $0.1091  \pm 0.0011$   &  $0.1094  \pm 0.0011$   &  $0.1091  \pm 0.0011$    \\[+1mm]
  $100\theta_\textrm{MC}$ & $1.04231 \pm0.00042$   &  $1.04233 \pm 0.00041$  &  $1.04228 \pm 0.00042$  &  $1.04235 \pm 0.00041$   \\[+1mm]
  $\tau$                  & $0.126   \pm0.018$     &  $0.107   \pm 0.017$    &  $0.130   \pm 0.018$    &  $0.107   \pm 0.015$     \\[+1mm]
  $\ln(10^{10} A_s)$      & $3.162   \pm0.036$     &  $3.121   \pm 0.034$    &  $3.169   \pm 0.035$    &  $3.121   \pm 0.030$     \\[+1mm]
  $\Omega_k$              & $-0.0257 \pm0.0091$    &  $-0.0133 \pm 0.0062$   &  $-0.0192 \pm 0.0060$   &  $-0.0132 \pm 0.0051$    \\[+1mm]
 \hline \\[-2mm]
  $H_0$ [km s$^{-1}$ Mpc$^{-1}$] & $61.5\pm2.9$    &  $66.0  \pm 2.5$        &  $63.6    \pm 2.2$      &  $66.0    \pm 2.0$      \\[+1mm]
  $\Omega_m$              & $0.355 \pm0.033$       &  $0.306  \pm 0.023$     &  $0.330   \pm 0.022$    &  $0.306   \pm 0.018$    \\[+1mm]
  $\sigma_8$              & $0.815  \pm 0.018$     &  $0.805 \pm 0.017$      &  $0.822   \pm 0.017$    &  $0.805   \pm 0.015$    \\[+0mm]
\end{tabular}
\end{ruledtabular}
\label{tab:para_SN}
\end{table*}

Table \ref{tab:para_flat_lcdm} lists the parameter constraints for the 
tilted spatially-flat and for the untilted nonflat $\Lambda$CDM inflation 
models, for the updated complete data set we use here. These constraints can be 
compared to those listed in the bottom right panels of Tables 5--8 of 
\citet{ParkRatra2018a} that were derived using the JLA SNIa data and the 
initial preprint value of the \citet{Ataetal2018} BAO distance measurement.
There are very small differences between the constraints derived using our 
previous and our updated full data sets. 

\begin{table*}
\caption{Tilted flat and untilted nonflat $\Lambda\textrm{CDM}$ model parameters constrained
with Planck, SN, BAO, $H(z)$, and $f\sigma_8$ data (mean and 68.3\% confidence limits).}
\begin{ruledtabular}
\begin{tabular}{lcc}
 \multicolumn{3}{c}{Tilted flat-$\Lambda\textrm{CDM}$ model} \\
 \hline \\[-2mm]
  Parameter                & TT+lowP+SN+BAO+$H(z)$+$f\sigma_8$  &  TT+lowP+lensing+SN+BAO+$H(z)$+$f\sigma_8$      \\[+0mm]
 \hline \\[-2mm]
  $\Omega_b h^2$           & $0.02233 \pm 0.00020$                      &  $0.02232 \pm 0.00019$     \\[+1mm]
  $\Omega_c h^2$           & $0.1178  \pm 0.0011$                       &  $0.1177  \pm 0.0011$      \\[+1mm]
  $100\theta_\textrm{MC}$  & $1.04104 \pm 0.00042$                      &  $1.04108 \pm 0.00041$     \\[+1mm]
  $\tau$                   & $0.070   \pm 0.017$                        &  $0.066   \pm 0.012$       \\[+1mm]
  $\ln(10^{10} A_s)$       & $3.069   \pm 0.033$                        &  $3.061   \pm 0.023$       \\[+1mm]
  $n_s$                    & $0.9693  \pm 0.0042$                       &  $0.9692  \pm 0.0043$      \\[+1mm]
 \hline \\[-2mm]
  $H_0$ [km s$^{-1}$ Mpc$^{-1}$] & $68.15   \pm 0.52$                   &  $68.19   \pm 0.50$        \\[+1mm]
  $\Omega_m$               & $0.3031  \pm 0.0067$                       &  $0.3025  \pm 0.0064$      \\[+1mm]
  $\sigma_8$               & $0.815   \pm 0.013$                        &  $0.8117  \pm 0.0088$      \\[+1mm]
  \hline \hline \\[-2mm]
 \multicolumn{3}{c}{Untilted nonflat $\Lambda\textrm{CDM}$ model} \\
 \hline \\[-2mm]
  Parameter                & TT+lowP+SN+BAO+$H(z)$+$f\sigma_8$  &  TT+lowP+lensing+SN+BAO+$H(z)$+$f\sigma_8$      \\[+0mm]
 \hline \\[-2mm]
  $\Omega_b h^2$           & $0.02307  \pm 0.00020$                     &  $0.02305  \pm 0.00019$    \\[+1mm]
  $\Omega_c h^2$           & $0.1094   \pm 0.0010$                      &  $0.1093   \pm 0.0010$    \\[+1mm]
  $100\theta_\textrm{MC}$  & $1.04225  \pm 0.00042$                     &  $1.04227  \pm 0.00041$    \\[+1mm]
  $\tau$                   & $0.121    \pm 0.016$                       &  $0.112    \pm 0.012$                   \\[+1mm]
  $\ln(10^{10} A_s)$       & $3.150    \pm 0.033$                       &  $3.132    \pm 0.022$      \\[+1mm]
  $\Omega_k$               & $-0.0083  \pm 0.0016$                      &  $-0.0083  \pm 0.0016$     \\[+1mm]
 \hline \\[-2mm]
  $H_0$ [km s$^{-1}$ Mpc$^{-1}$] & $67.96 \pm 0.62$                     &  $68.01 \pm 0.62$       \\[+1mm]
  $\Omega_m$               & $0.2882 \pm 0.0055$                        &  $0.2875 \pm 0.0055$     \\[+1mm]
  $\sigma_8$               & $0.820 \pm 0.014$                          &  $0.8121 \pm 0.0095$     \\[+0mm]
\end{tabular}
\end{ruledtabular}
\label{tab:para_flat_lcdm}
\end{table*}

\begin{figure*}
\centering
\mbox{\includegraphics[width=87mm]{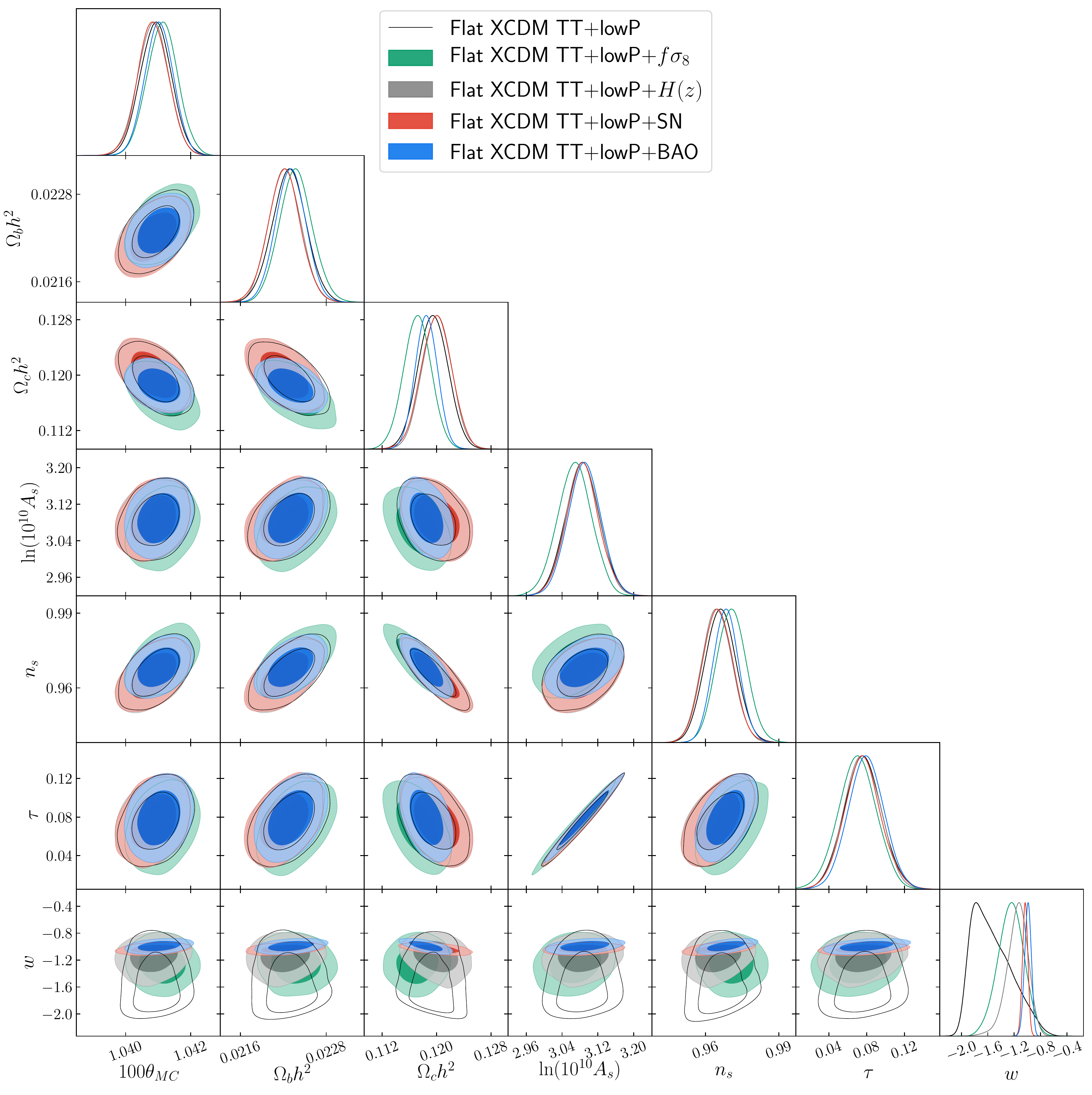}}
\mbox{\includegraphics[width=87mm]{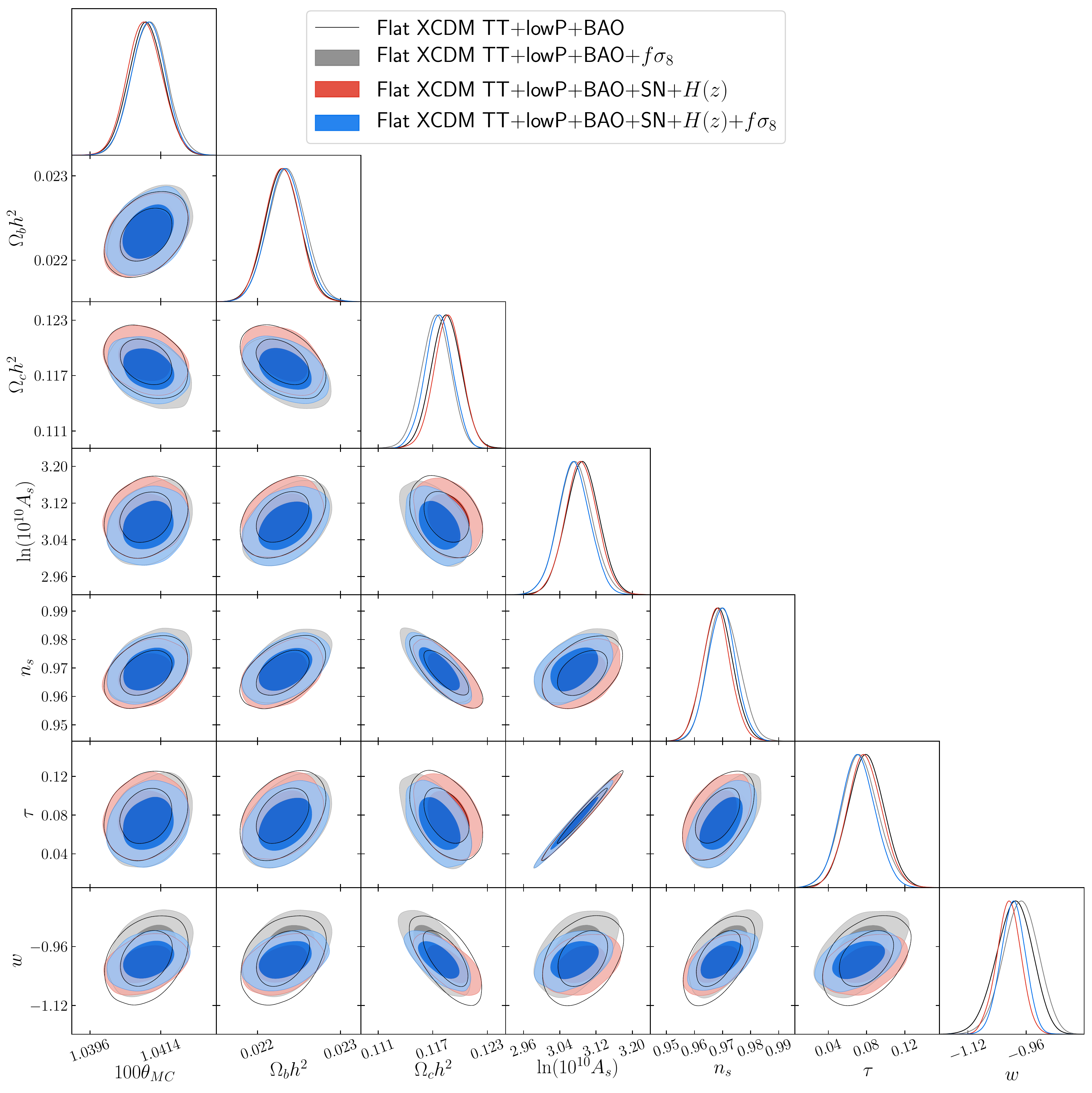}}
\caption{Likelihood distributions of the tilted flat-$\textrm{XCDM}$
model parameters constrained by Planck CMB TT + lowP, SN, BAO, $H(z)$,
and $f\sigma_8$ data.
Two-dimensional marginalized likelihood contours as well as
one-dimensional likelihoods are
shown for cases when each non-CMB measurement set is added to the 
Planck TT + lowP data
(left panel) and when the Hubble parameter, SN, growth rate data, and
the combination of them, are added to TT + lowP + BAO data (right panel).
For clarity the TT + lowP (left) and TT + lowP + BAO
(right panel) cases are shown as solid black curves.
}
\label{fig:para_flat}
\end{figure*}

\begin{figure*}
\centering
\mbox{\includegraphics[width=87mm]{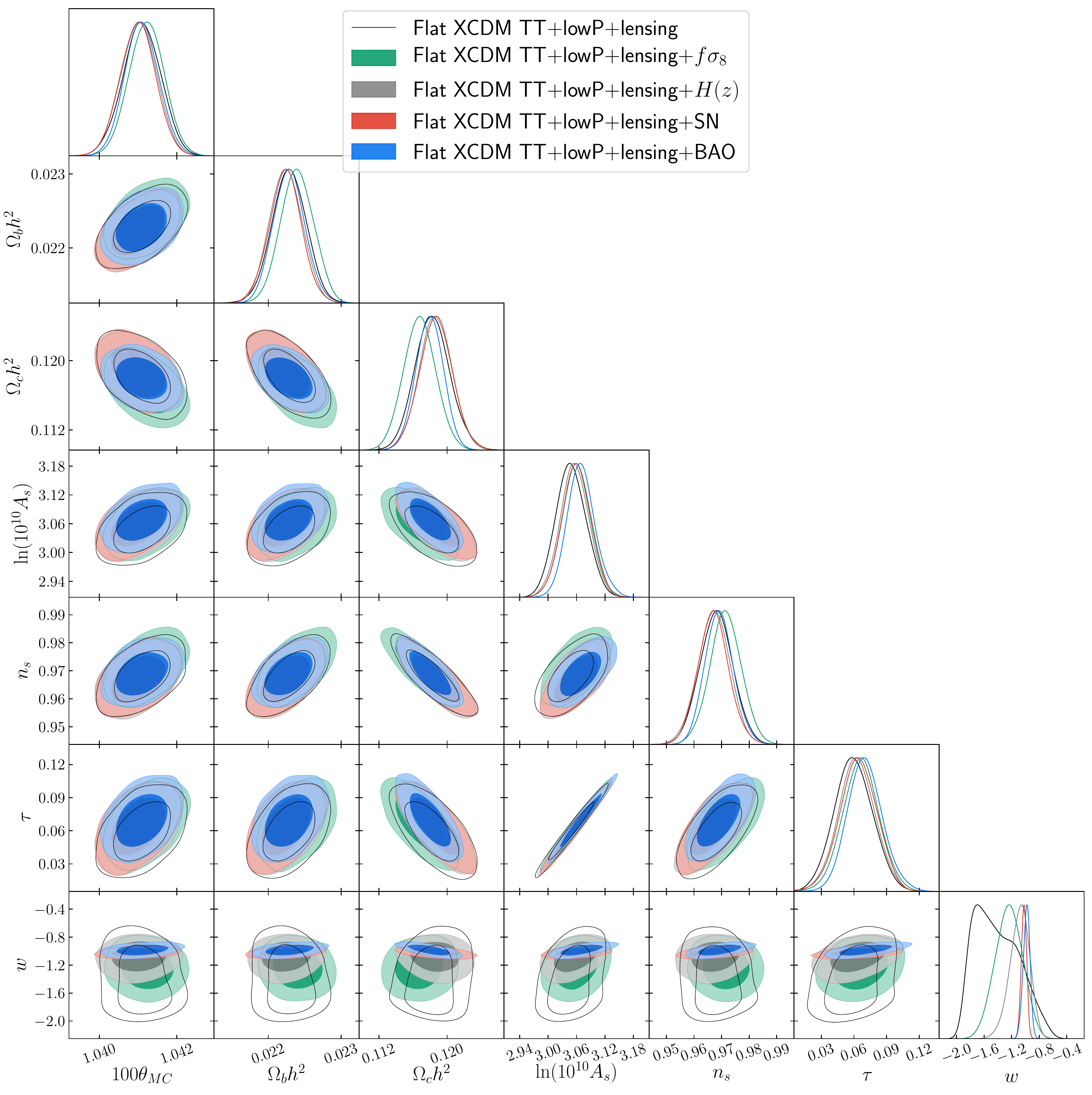}}
\mbox{\includegraphics[width=87mm]{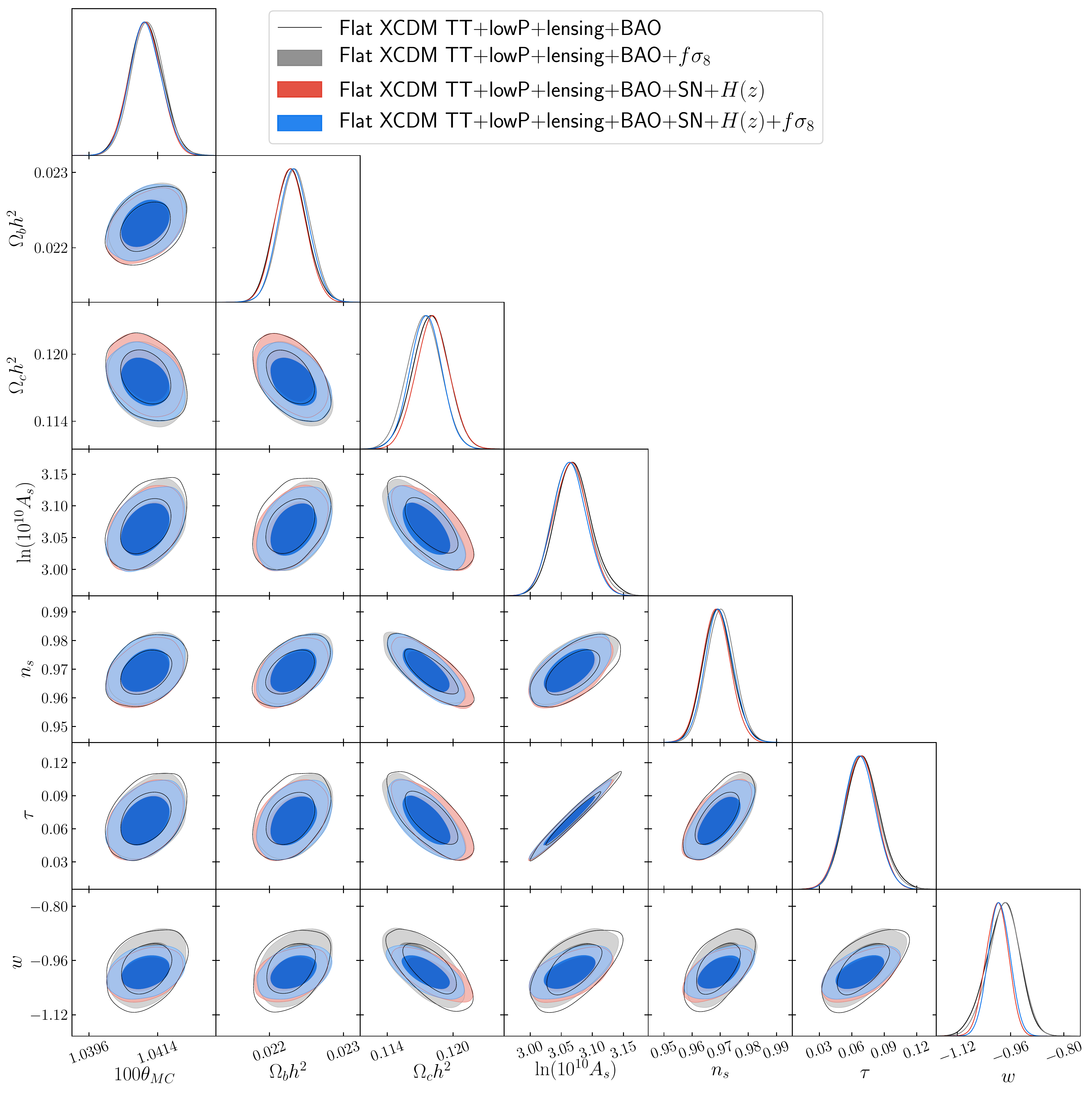}}
\caption{Same as Fig.\ \ref{fig:para_flat} but now also including
the Planck CMB lensing measurements.
}
\label{fig:para_flat_lensing}
\end{figure*}

\begin{figure*}
\centering
\mbox{\includegraphics[width=87mm]{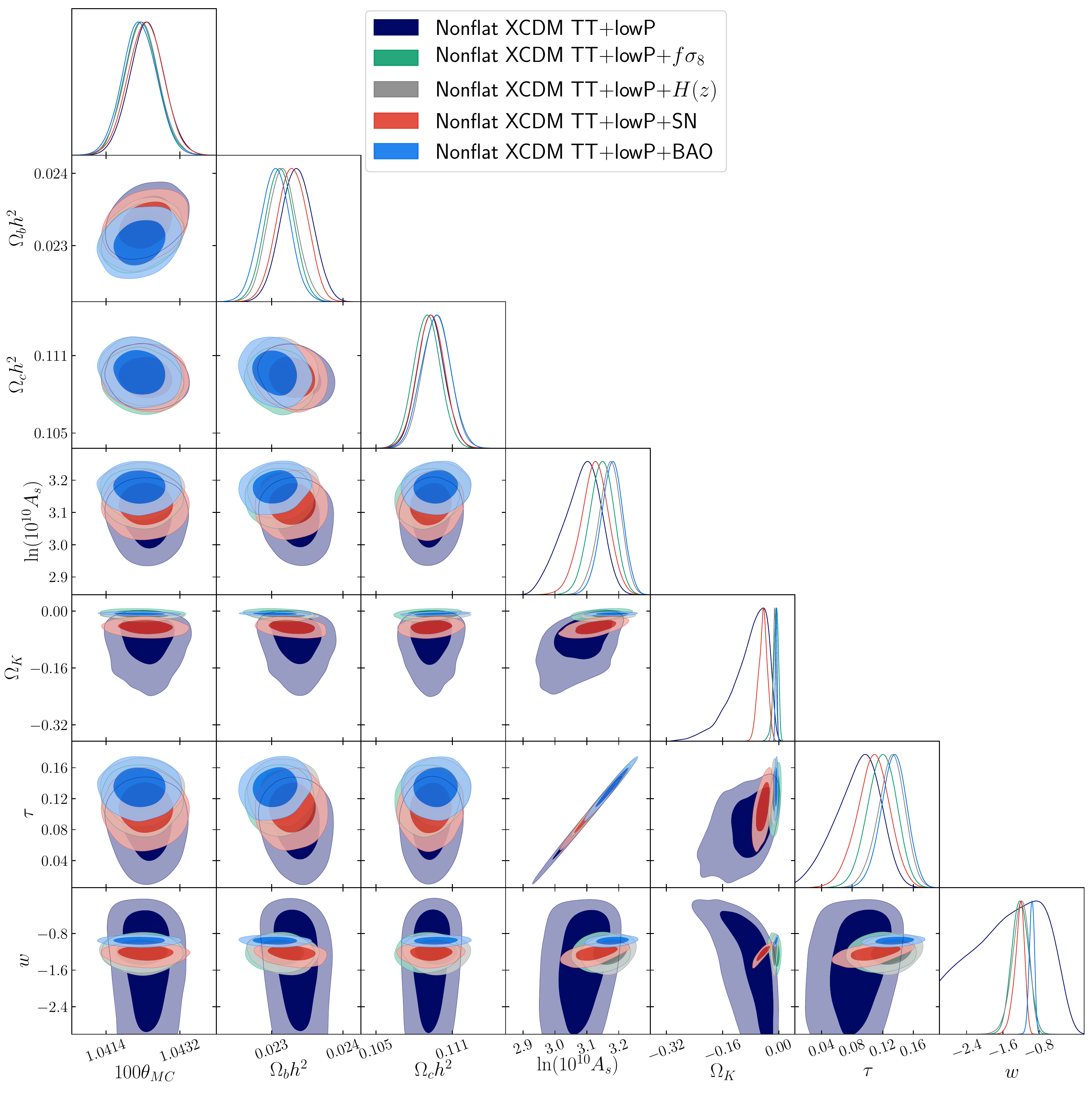}}
\mbox{\includegraphics[width=87mm]{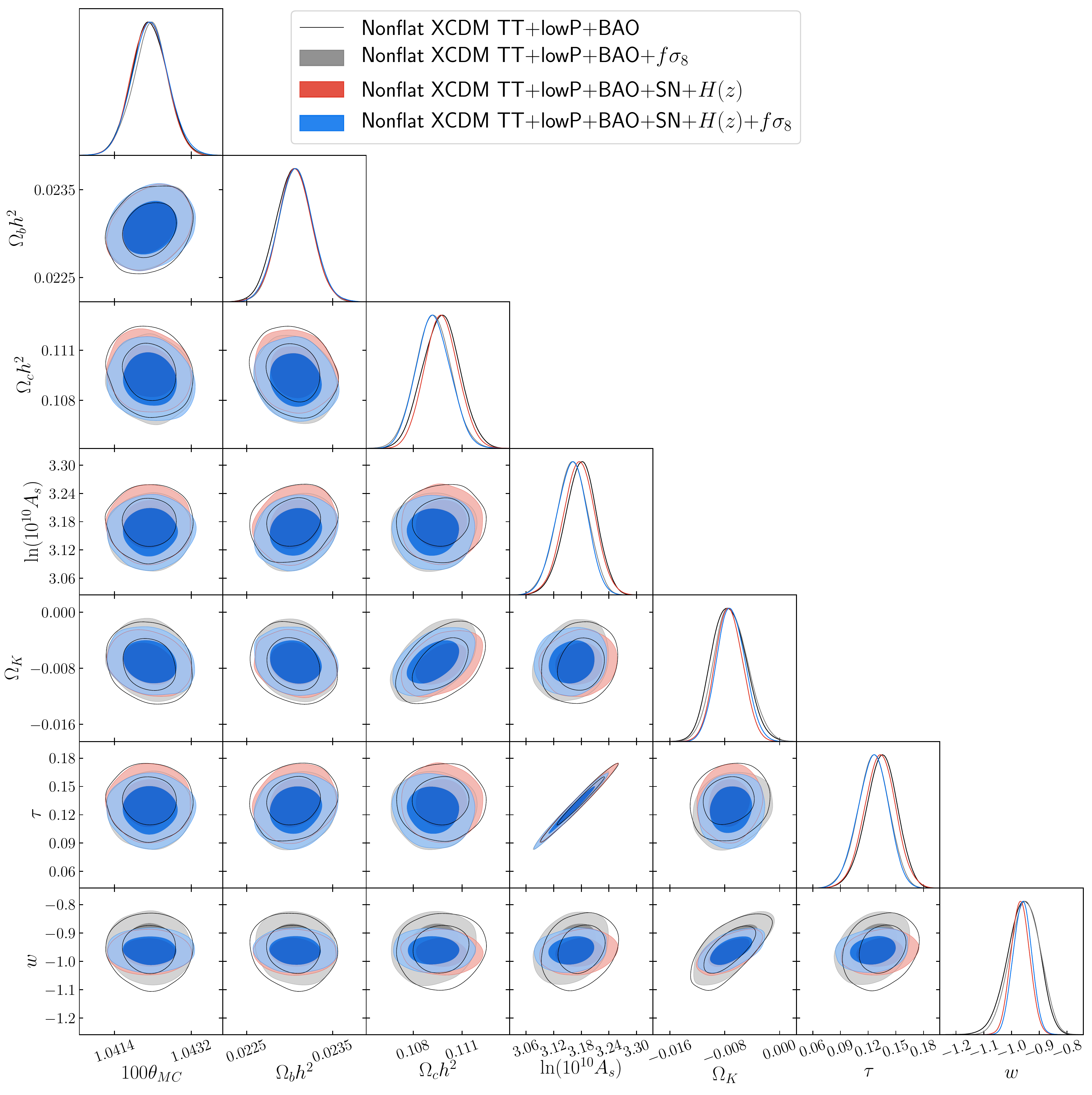}}
\caption{
Likelihood distributions of the untilted nonflat $\textrm{XCDM}$ model
parameters constrained by Planck CMB TT + lowP, SN, BAO, $H(z)$, and
$f\sigma_8$ data. Two-dimensional marginalized likelihood contours as 
well as one-dimensional likelihoods are shown for cases when each 
non-CMB measurement set is added to the Planck TT + lowP data
(left panel) and when the SN, Hubble parameter, growth rate data,
and the combination of them, are added to TT + lowP + BAO data
(right panel). For clarity, the result of TT + lowP + BAO
is shown as solid black curves in the right panel.
}
\label{fig:para_nonflat}
\end{figure*}

\begin{figure*}
\centering
\mbox{\includegraphics[width=87mm]{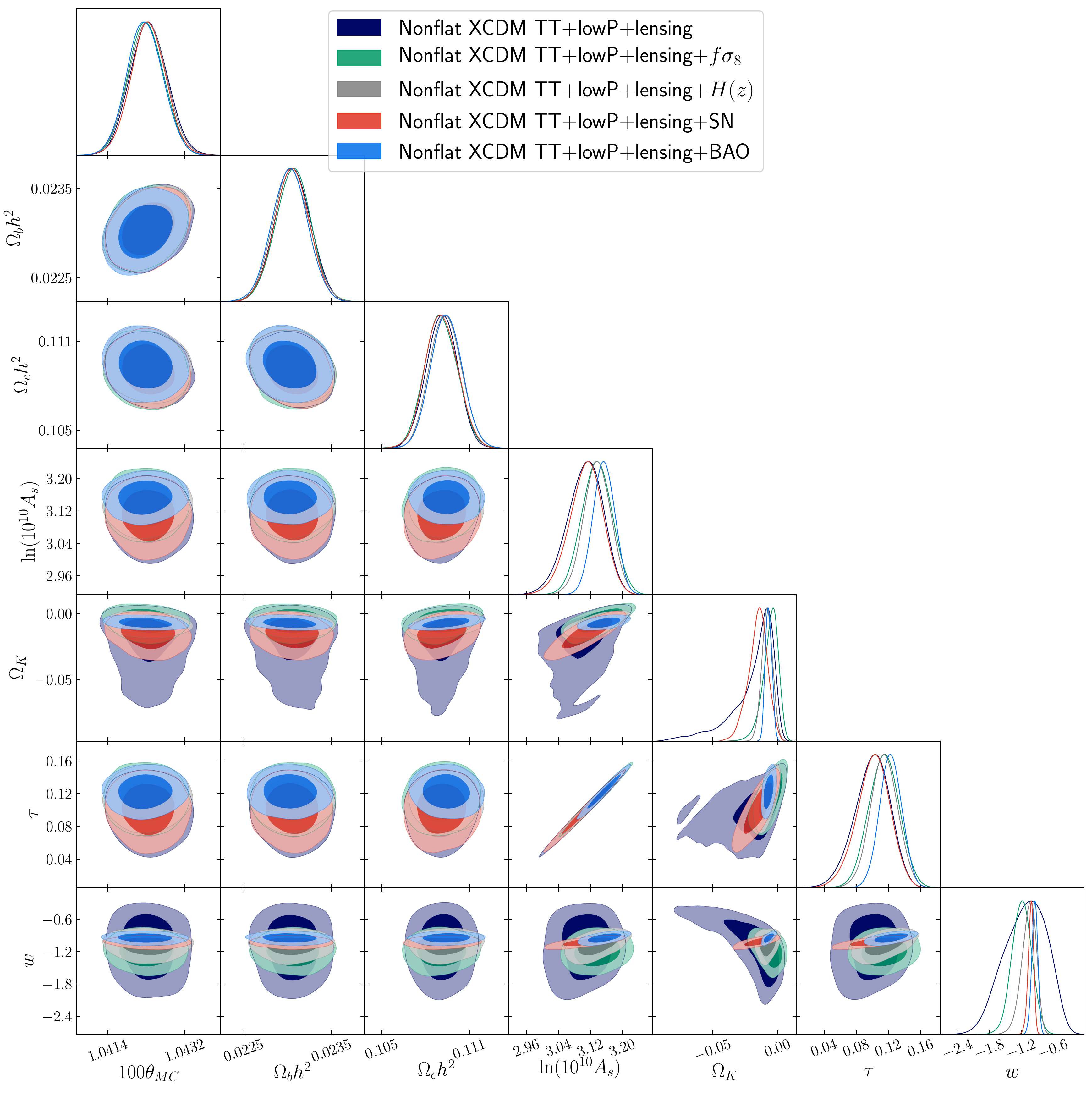}}
\mbox{\includegraphics[width=87mm]{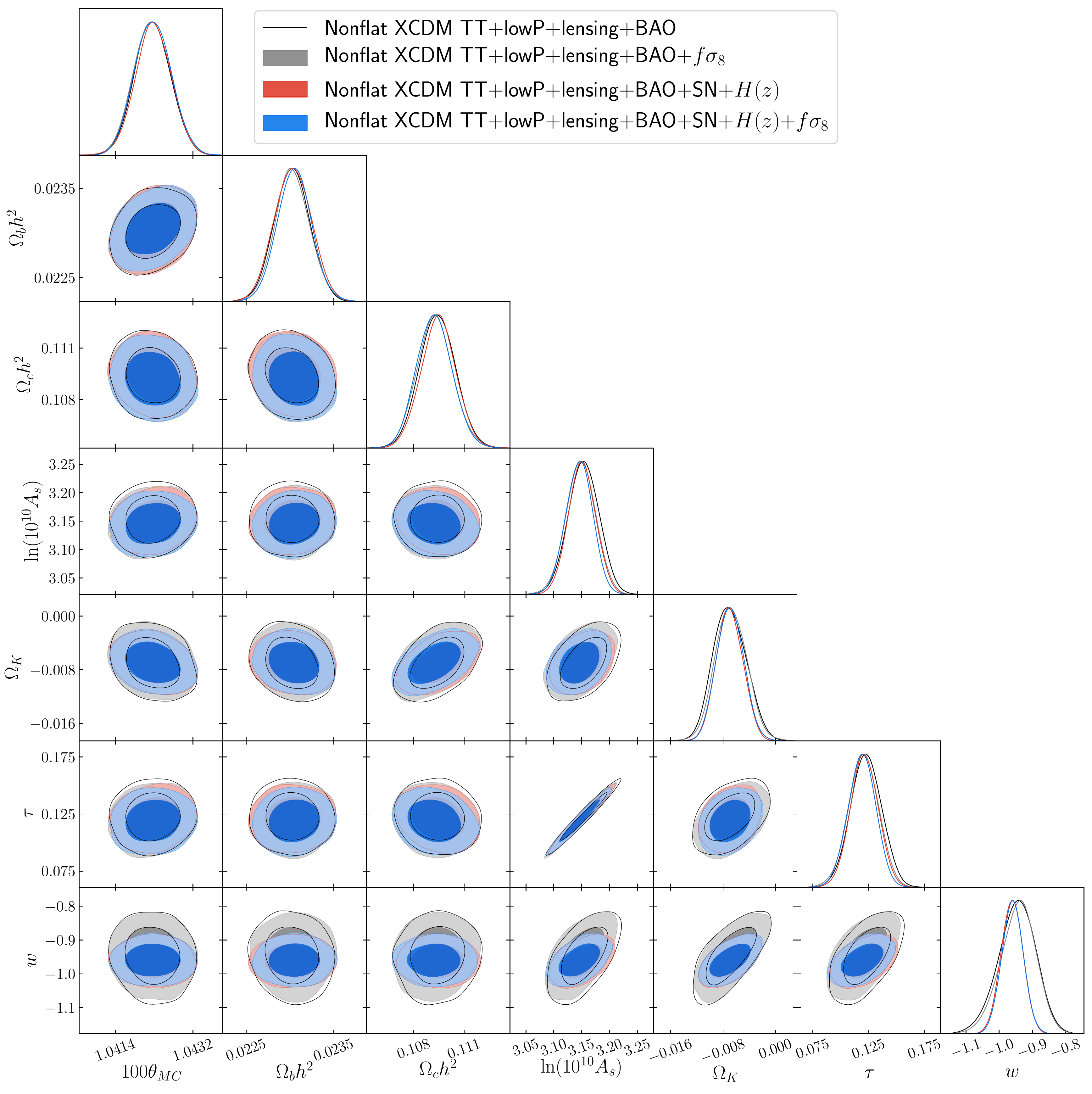}}
\caption{Same as Fig.\ \ref{fig:para_nonflat} but now also including
the Planck CMB lensing data.
}
\label{fig:para_nonflat_lensing}
\end{figure*}

Our results for the tilted flat and the nonflat XCDM parameterizations 
are presented in Figs.\ \ref{fig:para_flat}--\ref{fig:para_nonflat_lensing}
and Tables \ref{tab:para_flat}--\ref{tab:para_nonflat_lensing}.
In the plots we omit likelihood contours for
TT + lowP (+ lensing) + SN + BAO data (excluding or including
the Planck 2015 CMB lensing data) in both tilted flat and nonflat
XCDM cases because they are very similar to those for TT + lowP (+ lensing) 
+ SN + BAO + $H(z)$ data.
 
The entries for the tilted flat-XCDM parameterization in the TT + lowP 
panel of Table \ref{tab:para_flat} and in the TT + lowP + lensing panel 
in Table \ref{tab:para_flat_lensing} are very consistent with the corresponding 
Table 1 entries of \citet{Oobaetal2018d}, except for those for $w$, 
$H_0$, $\Omega_m$, and $\sigma_8$. This is because \citet{Oobaetal2018d}
use a flat prior non-zero over $0.2 \le h \le 1.3$ for $H_0$ while we use 
a flat prior non-zero over $0.2 \le h \le 1$.\footnote{Since the flat prior
on $h$ adopted here is the same as in the Planck team's
analyses, the parameters for the tilted flat-XCDM case constrained with
TT + lowP (+lensing) agree with the Planck results.
See base\_w\_plikHM\_TT\_lowTEB for TT + lowP data and
base\_w\_plikHM\_TT\_lowTEB\_post\_lensing for TT + lowP + lensing data \citep{PlanckParameterTables2015}.} 
The entries for the nonflat XCDM parameterization in the TT + lowP panel of Table 
\ref{tab:para_nonflat} and in the TT + lowP + lensing panel 
in Table \ref{tab:para_nonflat_lensing} agree well with the corresponding 
entries in Table 1 of \citet{Oobaetal2018b}. \citet{Oobaetal2018b} and
\citet{Oobaetal2018d} compute the $C_\ell$'s using CLASS \citep{Blasetal2011} 
and performed the MCMC analyses with Monte Python \citep{Audrenetal2013},
so it is reassuring that our results agree well with their results.
Our estimates of $w$, $\Omega_m$, and $H_0$ for the tilted flat-XCDM 
parameterization from the TT + lowP + SN data agree very well with the 
values presented in \citet{Scolnicetal2017}, $w=-1.031 \pm 0.040$, 
$\Omega_m=0.306 \pm 0.012$, and $H_0=68.335 \pm 1.098~\textrm{km}~\textrm{s}^{-1}~\textrm{Mpc}^{-1}$, which provides another reassuring check on our 
analyses.

From Tables \ref{tab:para_flat} and \ref{tab:para_flat_lensing} we see that, 
when they are added to the Planck CMB anisotropy observations, for the tilted 
flat-XCDM case, the BAO measurements mostly prove more restrictive 
than either the SN, $H(z)$, or $f \sigma_8$ data , except for $w$ where 
the SN data are more restrictive than the BAO data. However, when the 
CMB lensing data are included, Table \ref{tab:para_flat_lensing},
the CMB + SN limits on $w$, $H_0$, and $\sigma_8$ are more restrictive than
those from the CMB data combined with BAO, or $H(z)$, or $f\sigma_8$ 
measurements,
while all four non-CMB data sets used in conjunction with the CMB data provide
equally restrictive constraints on $\Omega_b h^2$ and $A_s$.\footnote{This
is not the case in the tilted flat-$\Lambda$CDM model, where for the data set
including the CMB lensing data, the CMB + BAO constraints on all parameters
are more restrictive than those determined by combining the CMB data with
either the SN, or $H(z)$, or $f\sigma_8$ measurements. For this model we
show only the CMB + SN constraints in Table \ref{tab:para_SN}.}
We note that our BAO compilation includes radial $H(z)$ measurements and 
the $f \sigma_8$ data of \citet{Alametal2017}. It is likely 
that if these are moved to the $H(z)$ and $f \sigma_8$ data sets, CMB and 
BAO, SN, $H(z)$, or $f\sigma_8$ constraints will all be about equally 
restrictive for the flat-XCDM parameterization.

The nonflat XCDM case is more interesting.
When CMB lensing data are included, Table \ref{tab:para_nonflat_lensing},
CMB data with either SN, or BAO, or $H(z)$, or $f \sigma_8$ data,
provide approximately equally restrictive constraints on $\Omega_b h^2$,
$\Omega_c h^2$, and $\theta_\textrm{MC}$, while CMB + BAO data provide
the tightest constraints on $\tau$, $A_s$, $\Omega_k$, $H_0$, $\Omega_m$, 
and $\sigma_8$, with CMB + SN setting tightest limits on 
$w$.\footnote{In the nonflat $\Lambda$CDM model (results 
mostly not shown here, except for CMB + SN shown in 
Table \ref{tab:para_SN}), $\Omega_b h^2$, $\Omega_c h^2$, and 
$\theta_\textrm{MC}$ are about equally well constrained by any of the 
four non-CMB data sets when used with the CMB (including lensing) data,
with CMB + BAO setting tighter limits on $\tau$, $A_s$, $\Omega_k$, $H_0$, 
$\Omega_m$, and $\sigma_8$.}

Focusing on the CMB TT + lowP + lensing measurements, Figs.\
\ref{fig:para_flat_lensing} and
\ref{fig:para_nonflat_lensing} and Tables \ref{tab:para_flat_lensing}
and \ref{tab:para_nonflat_lensing}, we see that adding each 
of the four non-CMB measurement sets at a time to the CMB data
(left triangle plots in both figures) produces four sets of
contours that are quite mutually consistent, as well as consistent with
the original CMB only contours, for both the tilted flat-XCDM
parameterization and for the untilted nonflat XCDM parameterization. 
It is reassuring that the four sets of non-CMB measurements do not push 
the CMB constraints 
in significantly different directions. This is also the 
case for the tilted flat-XCDM parameterization when the CMB lensing 
data are excluded (left triangle plot of Fig.\ \ref{fig:para_flat}).
However, in the untilted nonflat XCDM case excluding the lensing data
when any of the four sets of non-CMB observations are added to the CMB 
measurements (left triangle panel of Fig.\ \ref{fig:para_nonflat}),
they each push the results toward a smaller $|\Omega_k|$ 
(closer to spatially flat) and a slightly larger $\tau$ and $A_s$
and a smaller $\Omega_b h^2$ than what is favored by the CMB measurements alone,
though all five sets of constraint contours are mostly mutually 
consistent.
However, there is tension between the TT + lowP + SN and the 
TT + lowP + BAO contours in the $\Omega_k$--$w$ plane 
(Fig. \ref{fig:para_nonflat} left and Table \ref{tab:para_nonflat}),
where CMB + SN data give constraints on $\Omega_k$ and $w$ that deviate 
from the CMB + BAO data values by over $2\sigma$, with $w$ also 
deviating from the cosmological constant ($w=-1$) by over $2\sigma$
for the CMB + SN case and $\Omega_k$ differing from $0$ by more than
3$\sigma$ in both cases. 

While adding the BAO data to the CMB data usually results in the biggest
difference, the other three non-CMB sets of data also contribute. 
Focusing on TT + lowP + lensing data, we see from Table 
\ref{tab:para_flat_lensing} for the tilted flat-XCDM parameterization that
the BAO data tightly constrains model parameters, especially
$\Omega_c h^2$, while the $f\sigma_8$ measurements push
$\Omega_b h^2$ and $n_s$ to larger values and push $\Omega_c h^2$ to a 
smaller value. In this case $H_0$ is the parameter whose error bar is
decreased the most for the full combination of data relative to the CMB + SN 
data combination, followed by the $\Omega_m$ error bar reduction relative to 
CMB + BAO data combination. For the untilted nonflat XCDM case, from Table 
\ref{tab:para_nonflat_lensing}, the error bars that shrink the most when 
CMB (including lensing) data are used in conjunction with the four 
non-CMB data sets are those on $w$ (relative to the CMB + SN case)
and $H_0$ and $\Omega_m$ (relative to the CMB + BAO combination).

Continuing to focus on the TT + lowP + lensing data, Tables   
\ref{tab:para_flat_lensing} and \ref{tab:para_nonflat_lensing}, we see that 
for the tilted flat-XCDM parameterization, adding the four sets of non-CMB data 
to the mix most influences $\sigma_8$, $w$, and $\Omega_m$, with the $\sigma_8$
central value moving down by 1.3$\sigma$ and the $w$ and $\Omega_m$ central
values moving up by 1.3$\sigma$ and 1.2$\sigma$, all of the CMB data alone
error bars; $\theta_\textrm{MC}$ is hardly affected by adding
the four non-CMB data sets, changing by only 0.042$\sigma$. The 
situation for the nonflat XCDM parameterization is a little less
dramatic, with $\ln (10^{10} A_s)$, and $\tau$ central values increasing
by 0.91$\sigma$ and 0.86$\sigma$ of the CMB data alone error bars, and 
$\Omega_k$ moving closer to flatness by 0.71$\sigma$; the $\Omega_b h^2$
central value does not change in this case.

\begin{figure*}
\centering
\mbox{\includegraphics[width=65mm]{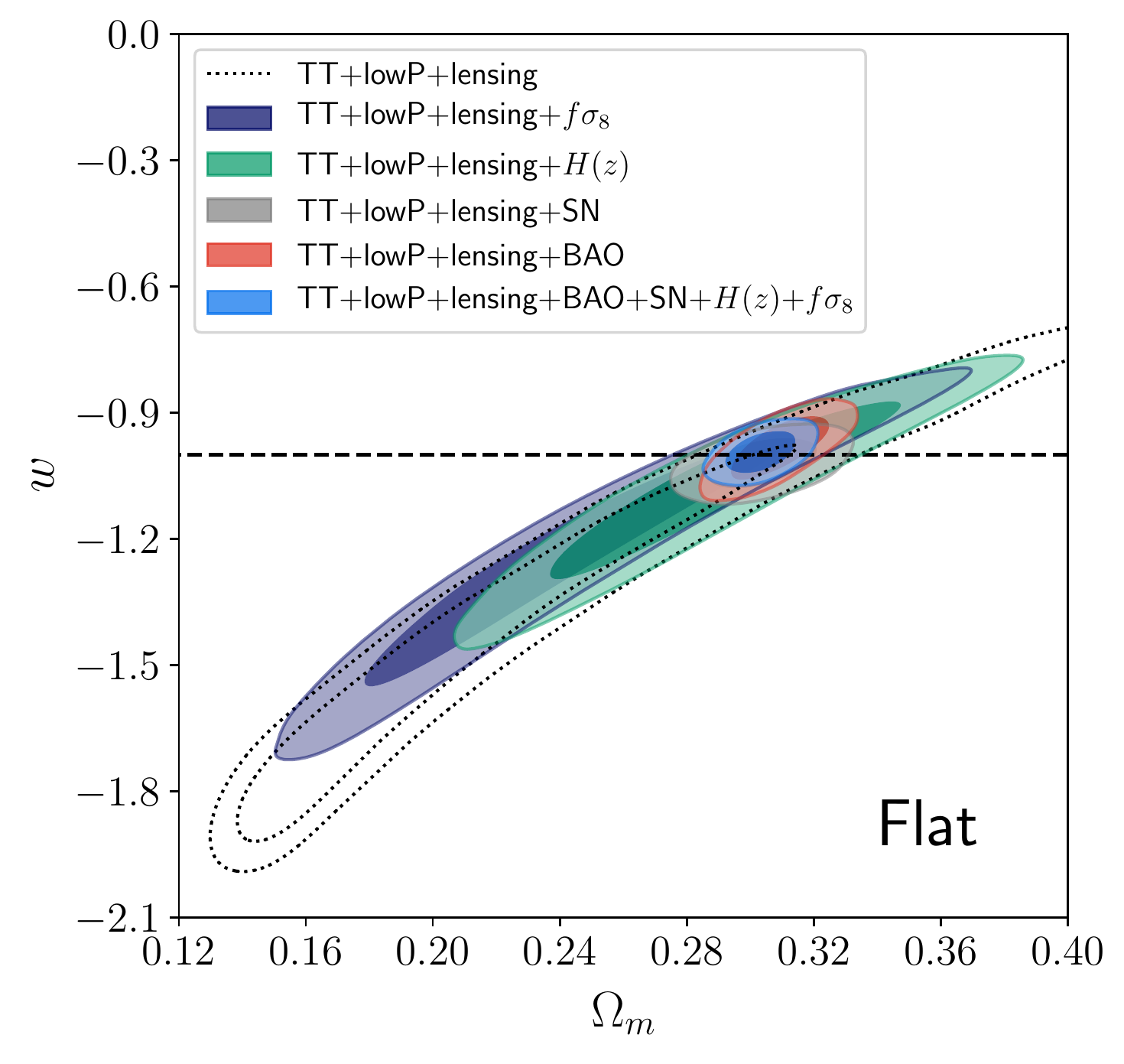}}
\mbox{\includegraphics[width=68.4mm]{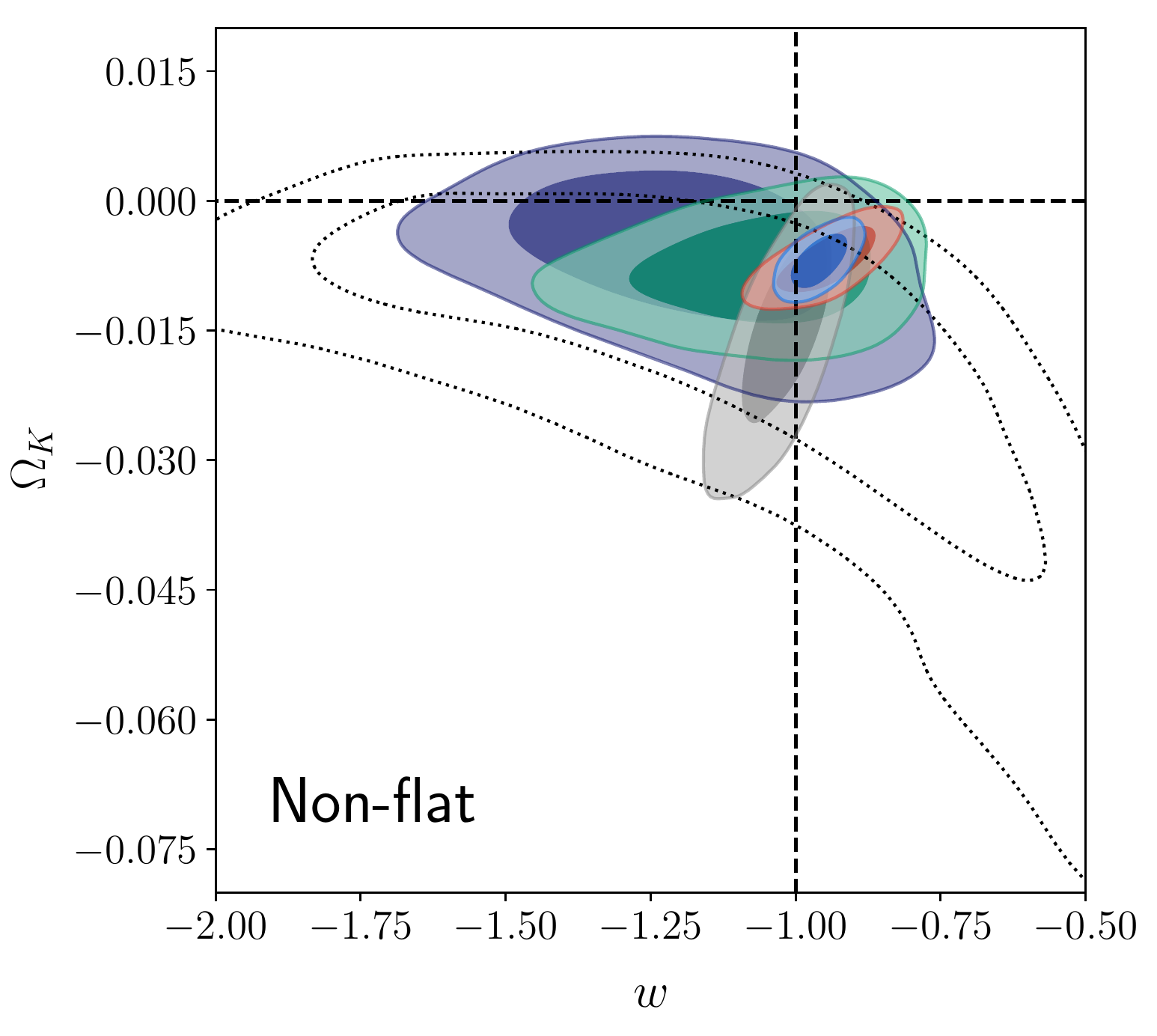}}
\caption{1$\sigma$ and 2$\sigma$ likelihood contours in the $\Omega_m$--$w$ plane
for the tilted flat-XCDM parameterization (left panel) and in the $w$--$\Omega_k$
plane for the untilted nonflat $\textrm{XCDM}$ parameterization (right panel),
constrained by Planck CMB TT + lowP + lensing and non-CMB data sets.
The horizontal and vertical dashed lines indicate $w=-1$ (the cosmological constant)
or $\Omega_k=0$. Contours in both panels follow the color scheme shown in the left panel.
}
\label{fig:omm_w_omk}
\end{figure*}

Figure \ref{fig:omm_w_omk} shows marginalized likelihood contours in 
the $\Omega_m$--$w$ plane for the tilted flat-XCDM parameterization
and in the $w$--$\Omega_k$ plane for the untilted nonflat XCDM case. For 
CMB TT + lowP 
+ lensing data combined with the non-CMB data sets, the flat-XCDM 
parameterization prefers $w=-1$, favoring the cosmological constant 
as dark energy. On the other hand, the nonflat XCDM parameterization, 
when constrained by the full data, prefers closed spatial hypersurfaces 
and a dark energy equation of state parameter $w > -1$.

More precisely, including the four non-CMB sets of measurements in the mix, 
we find 
in the tilted flat-XCDM parameterization (bottom right panel of Table 
\ref{tab:para_flat_lensing}) that $w = -0.994 \pm 0.033$, which is more tightly 
restricted to $ w = -1$ and the cosmological constant than the original
\citet{Oobaetal2018d} finding of $w = -1.03 \pm 0.07$ (the last column of their
Table 1).\footnote{These results differ from those of earlier approximate 
analyses, based on less and less reliable data, that indicated evidence 
for $w$ deviating from $-1$ by more than 3$\sigma$ \citep{Solaetal2017a, Solaetal2018, Solaetal2017b, Solaetal2017c, Solaetal2017d, GomezValentSola2017, GomezValentSola2018}.}

On the other hand, and perhaps the most striking consequence of adding 
the four non-CMB data sets to the mix here, is the significant 
strengthening of the support for nonflatness in the untilted nonflat XCDM 
case, with it increasing to $\Omega_k = - 0.0069 \pm 0.0020$, more than 
3.4$\sigma$ away from flatness now, for the full data combination in the 
bottom right panel of Table \ref{tab:para_nonflat_lensing}, compared to 
the 1.1$\sigma$ from flatness for the CMB only case. That is 
now accompanied for the first time by mild evidence favoring dynamical 
dark energy with $w = - 0.960 \pm 0.032$ that is more than 1.2$\sigma$
away from the cosmological constant. These results are consistent with but
strengthen those of \cite{Oobaetal2018b} who found $\Omega_k = - 0.008 \pm 
0.003$ and  $w = - 1.00 \pm 0.10$ from Planck 2015 CMB anisotropy data 
in combination with a few BAO distance measurements. The stronger results 
here are driven in part by each of the four non-CMB data sets. CMB + BAO,
CMB + SN, and CMB + $H(z)$ favor negative $\Omega_k$ values 2.8$\sigma$, 
2.0$\sigma$, and 1.9$\sigma$ away from flat, while CMB + $f\sigma_8$ and 
CMB + BAO favor $w$ values that are 1.1$\sigma$ more negative and 1$\sigma$ 
less negative than $w = -1$. On the other hand, CMB + $f\sigma_8$ data are 
consistent with a flat model and CMB + SN and CMB + $H(z)$ are consistent
with the cosmological constant and $w = -1$. In favoring a closed model 
with $w$ less negative than $-1$, the BAO data play the most important 
role amongst the four non-CMB data sets.

\begin{table*}
\caption{Tilted flat-$\textrm{XCDM}$ model parameters constrained
with Planck TT + lowP, SN, BAO, $H(z)$, and $f\sigma_8$ data (mean and 68.3\% confidence limits).}
\begin{ruledtabular}
\begin{tabular}{lccc}
  Parameter                & TT+lowP                & TT+lowP+SN               &  TT+lowP+BAO      \\[+0mm]
 \hline \\[-2mm]
  $\Omega_b h^2$           & $0.02228 \pm 0.00023$  & $0.02221 \pm 0.00023$    &  $0.02231  \pm 0.00021$    \\[+1mm]
  $\Omega_c h^2$           & $0.1195  \pm 0.0022$   & $0.1200  \pm 0.0022$     &  $0.1185   \pm 0.0016$       \\[+1mm]
  $100\theta_\textrm{MC}$  & $1.04093 \pm 0.00048$  & $1.04085 \pm 0.00047$    &  $1.04103  \pm 0.00044$     \\[+1mm]
  $\tau$                   & $0.076 \pm 0.020$      & $0.076   \pm 0.019$      &  $0.079    \pm 0.019$         \\[+1mm]
  $\ln(10^{10} A_s)$       & $3.086 \pm 0.037$      & $3.086   \pm 0.037$      &  $3.090    \pm 0.036$        \\[+1mm]
  $n_s$                    & $0.9662 \pm 0.0063$    & $0.9651  \pm 0.0061$     &  $0.9684   \pm 0.0052$     \\[+1mm]
  $w$                      & $-1.53 \pm 0.30$       & $-1.034  \pm 0.040$      &  $-0.993   \pm 0.050$     \\[+1mm]
 \hline \\[-2mm]
  $H_0$ [km s$^{-1}$ Mpc$^{-1}$] & $>63.5~(95.4\%~\textrm{C.L.})$ & $68.2  \pm 1.1$        &  $67.6     \pm 1.2$          \\[+1mm]
  $\Omega_m$               & $0.207 \pm 0.057$      & $0.307 \pm 0.012$        &  $0.309    \pm 0.010$      \\[+1mm]
  $\sigma_8$               & $0.977 \pm 0.084$      & $0.838 \pm 0.019$        &  $0.824    \pm 0.019$        \\[+1mm]
  \hline \hline \\[-2mm]
    Parameter              & TT+lowP+$H(z)$         &  TT+lowP+SN+BAO  &  TT+lowP+SN+BAO+$H(z)$  \\[+0mm]
 \hline \\[-2mm]
  $\Omega_b h^2$           & $0.02222 \pm 0.00022$  &  $0.02229  \pm 0.00021$  &  $0.02230 \pm 0.00020$    \\[+1mm]
  $\Omega_c h^2$           & $0.1200  \pm 0.0021$   &  $0.1186   \pm 0.0015$   &  $0.1187  \pm 0.0015$     \\[+1mm]
  $100\theta_\textrm{MC}$  & $1.04085 \pm 0.00047$  &  $1.04098  \pm 0.00044$  &  $1.04100 \pm 0.00044$    \\[+1mm]
  $\tau$                   & $0.075 \pm 0.019$      &  $0.078    \pm 0.018$    &  $0.078   \pm 0.018$      \\[+1mm]
  $\ln(10^{10} A_s)$       & $3.086 \pm 0.037$      &  $3.088    \pm 0.035$    &  $3.088   \pm 0.036$      \\[+1mm]
  $n_s$                    & $0.9648 \pm 0.0061$    &  $0.9677   \pm 0.0051$   &  $0.9679  \pm 0.0050$     \\[+1mm]
  $w$                      & $-1.15 \pm 0.16$       &  $-1.006   \pm 0.034$    &  $-1.007  \pm 0.034$     \\[+1mm]
 \hline \\[-2mm]
  $H_0$ [km s$^{-1}$ Mpc$^{-1}$] & $71.9 \pm 4.8$   & $67.97    \pm 0.77$      &  $68.00   \pm 0.77$       \\[+1mm]
  $\Omega_m$               & $0.280 \pm 0.037$      & $0.3066   \pm 0.0076$    &  $0.3064  \pm 0.0075$     \\[+1mm]
  $\sigma_8$               & $0.872 \pm 0.047$      & $0.827    \pm 0.016$     &  $0.828   \pm 0.016$      \\[+1mm]
    \hline \hline \\[-2mm]
    Parameter              & TT+lowP+$f\sigma_8$    & TT+lowP+BAO+$f\sigma_8$  &  TT+lowP+SN+BAO+$H(z)$+$f\sigma_8$  \\[+0mm]
 \hline \\[-2mm]
  $\Omega_b h^2$           & $0.02238 \pm 0.00023$  & $0.02235 \pm 0.00022$    &  $0.02234 \pm 0.00021$    \\[+1mm]
  $\Omega_c h^2$           & $0.1172  \pm 0.0020$   & $0.1173  \pm 0.0016$     &  $0.1177  \pm 0.0015$     \\[+1mm]
  $100\theta_\textrm{MC}$  & $1.04113 \pm 0.00047$  & $1.04111 \pm 0.00044$    &  $1.04109 \pm 0.00043$    \\[+1mm]
  $\tau$                   & $0.070 \pm 0.020$      & $0.073   \pm 0.020$      &  $0.071   \pm 0.018$      \\[+1mm]
  $\ln(10^{10} A_s)$       & $3.068 \pm 0.038$      & $3.075   \pm 0.037$      &  $3.071   \pm 0.035$      \\[+1mm]
  $n_s$                    & $0.9706 \pm 0.0061$    & $0.9704  \pm 0.0054$     &  $0.9697  \pm 0.0050$     \\[+1mm]
  $w$                      & $-1.25 \pm 0.20$       & $-0.975  \pm 0.048$      &  $-0.996  \pm 0.034$     \\[+1mm]
 \hline \\[-2mm]
  $H_0$ [km s$^{-1}$ Mpc$^{-1}$] & $76.8 \pm 6.9$   & $67.5  \pm 1.1$          &  $68.07   \pm 0.78$       \\[+1mm]
  $\Omega_m$               & $0.244 \pm 0.045$      & $0.3079 \pm 0.0098$      &  $0.3037  \pm 0.0075$     \\[+1mm]
  $\sigma_8$               & $0.885 \pm 0.056$      & $0.809 \pm 0.017$        &  $0.814   \pm 0.015$       \\[+0mm]
\end{tabular}
\end{ruledtabular}
\label{tab:para_flat}
\end{table*}

For the full data combination (including CMB lensing data) in 
Tables \ref{tab:para_flat_lensing} and \ref{tab:para_nonflat_lensing}, 
$H_0$ values measured using the tilted flat-XCDM and the nonflat 
XCDM parameterizations, $68.06 \pm 0.77$ and $67.45 \pm 0.75$ 
km s$^{-1}$ Mpc$^{-1}$, are consistent with each other to within
0.57$\sigma$ (of the quadrature sum of the two error bars).\footnote{Potential
systematic errors that might affect the value of $H_0$, ignored here, have 
been discussed by \citet{Addisonetal2016} and \citet{PlanckCollaboration2017}.}
These values are very compatible with the median statistics estimate
$H_0=68\pm2.8$ km s$^{-1}$ Mpc$^{-1}$ 
\citep{ChenRatra2011a}, which agrees with earlier median statistics 
measurements \citep{Gottetal2001, Chenetal2003}. Other recent measurements
of $H_0$ are also very compatible with these estimates \citep{Aubourgetal2015,PlanckCollaboration2016,SemizCamlibel2015,LHuillierShafieloo2017,Chenetal2017,Lukovicetal2016,Wangetal2017,LinIshak2017,DESCollaboration2017,Yuetal2018,Haridasuetal2018,Zhangetal2018,GomezValentAmendola2018}, but, as well known, these 
estimates are lower than the local expansion rate measurement of 
$H_0 = 73.48\pm1.66$ km s$^{-1}$ Mpc$^{-1}$
\citep{Riessetal2018}.\footnote{This local measurement is 2.9$\sigma$ (3.3$\sigma$), of the quadrature sum of both error bars, higher than $H_0$ measured 
here using the tilted flat-XCDM (untilted nonflat XCDM) parameterization. Other
local expansion rate estimates find slightly lower 
$H_0$'s with larger error bars
\citep{Rigaultetal2015, Zhangetal2017b, Dhawanetal2018, FernandezArenasetal2018}.}

In our analyses here, $H_0$ and $\sigma_8$ (see below) are the only 
cosmological parameters that are measured in an almost cosmological model 
(tilt and spatial curvature) independent way. Measurements of other 
cosmological parameters determined using the two XCDM parameterizations 
differ more significantly. More precisely, measurements determined using
the full data set (including CMB lensing) of $w$, $\Omega_m$, 
$\theta_\textrm{MC}$, $\ln(10^{10} A_s)$, $\Omega_b h^2$, $\tau$, and  
$\Omega_c h^2$, differ by 0.74$\sigma$, 1.1$\sigma$, 2.0$\sigma$, 
2.3$\sigma$, 2.5$\sigma$, 2.7$\sigma$, and 4.8$\sigma$ (of the quadrature 
sum of both error bars). For some of these parameters, especially 
$\Omega_c h^2$ as well as probably $\tau$ and $\Omega_b h^2$, the cosmological
model dependence of the measurement creates a much larger uncertainty than
does the statistical error in a given cosmological model. This effect 
was first noticed in a comparison between measurements made using the    
tilted flat-$\Lambda$CDM and the untilted nonflat $\Lambda$CDM inflation models 
\citep{ParkRatra2018a}. From Tables \ref{tab:para_flat_lensing} and 
\ref{tab:para_nonflat_lensing}, for the full data compilation (including 
CMB lensing), we find in the tilted flat-XCDM (nonflat XCDM) case 
$0.038 \le \tau \le 0.098$ ($0.095 \le \tau \le 0.143$) and
$0.02191 \le \Omega_b h^2 \le 0.02275$ 
($0.02265 \le \Omega_b h^2 \le 0.02345$) at 2$\sigma$, which are almost 
completely disjoint. It is not yet possible to measure $\Omega_c h^2$, $\tau$, or  
$\Omega_b h^2$ (and possibly some other cosmological parameters as well) 
in a model independent way by using cosmological data.


\begin{table*}
\caption{Tilted flat-$\textrm{XCDM}$ model parameters constrained with Planck TT + lowP + lensing, SN, BAO, $H(z)$, and $f\sigma_8$ data (mean and 68.3\% confidence limits).}
\begin{ruledtabular}
\begin{tabular}{lccc}
  Parameter                & TT+lowP+lensing        & TT+lowP+lensing+SN  &  TT+lowP+lensing+BAO   \\[+0mm]
 \hline \\[-2mm]
  $\Omega_b h^2$           & $0.02229 \pm 0.00023$  & $0.02223 \pm 0.00022$  &  $0.02229  \pm 0.00022$    \\[+1mm]
  $\Omega_c h^2$           & $0.1183  \pm 0.0021$   & $0.1187  \pm 0.0019$   &  $0.1179   \pm 0.0016$       \\[+1mm]
  $100\theta_\textrm{MC}$  & $1.04110 \pm 0.00048$  & $1.04099 \pm 0.00045$  &  $1.04110  \pm 0.00044$     \\[+1mm]
  $\tau$                   & $0.060 \pm 0.017$      & $0.064   \pm 0.017$    &  $0.070    \pm 0.016$         \\[+1mm]
  $\ln(10^{10} A_s)$       & $3.048 \pm 0.032$      & $3.060   \pm 0.030$    &  $3.070    \pm 0.030$        \\[+1mm]
  $n_s$                    & $0.9681 \pm 0.0060$    & $0.9671  \pm 0.0056$   &  $0.9692   \pm 0.0052$     \\[+1mm]
  $w$                      & $-1.41 \pm 0.32$       & $-1.020  \pm 0.039$    &  $-0.984   \pm 0.050$     \\[+1mm]
 \hline \\[-2mm]
  $H_0$ [km s$^{-1}$ Mpc$^{-1}$] & $>60.1~(95.4\%~\textrm{C.L.})$ & $68.3  \pm 1.1$   &  $67.6  \pm 1.2$     \\[+1mm]
  $\Omega_m$               & $0.223 \pm 0.068$      & $0.303 \pm 0.012$      &  $0.309    \pm 0.010$      \\[+1mm]
  $\sigma_8$               & $0.924 \pm 0.085$      & $0.820 \pm 0.013$      &  $0.811    \pm 0.014$        \\[+1mm]
  \hline \hline \\[-2mm]
    Parameter              & TT+lowP+lensing+$H(z)$ &  TT+lowP+lensing+SN+BAO  &  TT+lowP+lensing+SN+BAO+$H(z)$  \\[+0mm]
 \hline \\[-2mm]
  $\Omega_b h^2$           & $0.02224 \pm 0.00022$  &  $0.02227  \pm 0.00021$ &  $0.02228 \pm 0.00021$    \\[+1mm]
  $\Omega_c h^2$           & $0.1187  \pm 0.0019$   &  $0.1182   \pm 0.0015$   &  $0.1181  \pm 0.0015$     \\[+1mm]
  $100\theta_\textrm{MC}$  & $1.04101 \pm 0.00046$  &  $1.04105  \pm 0.00043$  &  $1.04107 \pm 0.00042$    \\[+1mm]
  $\tau$                   & $0.063 \pm 0.017$      &  $0.067    \pm 0.015$    &  $0.068   \pm 0.015$      \\[+1mm]
  $\ln(10^{10} A_s)$       & $3.056 \pm 0.030$      &  $3.064    \pm 0.028$    &  $3.065   \pm 0.028$      \\[+1mm]
  $n_s$                    & $0.9672 \pm 0.0058$    &  $0.9681   \pm 0.0050$   &  $0.9686  \pm 0.0050$     \\[+1mm]
  $w$                      & $-1.08  \pm 0.14$      &  $-1.001   \pm 0.0034$  &  $-1.000  \pm 0.033$     \\[+1mm]
 \hline \\[-2mm]
  $H_0$ [km s$^{-1}$ Mpc$^{-1}$]  & $70.4 \pm 4.5$  & $67.97    \pm 0.77$      &  $70.00   \pm 0.76$       \\[+1mm]
  $\Omega_m$               & $0.289 \pm 0.037$      & $0.3057   \pm 0.0074$    &  $0.3052  \pm 0.0073$     \\[+1mm]
  $\sigma_8$               & $0.836 \pm 0.038$      & $0.815    \pm 0.011$     &  $0.815   \pm 0.011$      \\[+1mm]
  \hline  \hline \\[-2mm]
    Parameter              & TT+lowP+lensing+$f\sigma_8$  & TT+lowP+lensing+BAO+$f\sigma_8$ &  TT+lowP+lensing+SN+BAO+$H(z)$+$f\sigma_8$  \\[+0mm]
 \hline \\[-2mm]
  $\Omega_b h^2$           & $0.02240 \pm 0.00022$  & $0.02234 \pm 0.00021$    &  $0.02233 \pm 0.00021$    \\[+1mm]
  $\Omega_c h^2$           & $0.1168  \pm 0.0019$   & $0.1173  \pm 0.0015$     &  $0.1175  \pm 0.0014$     \\[+1mm]
  $100\theta_\textrm{MC}$  & $1.04122 \pm 0.00045$  & $1.04114 \pm 0.00042$    &  $1.04108 \pm 0.00042$    \\[+1mm]
  $\tau$                   & $0.066 \pm 0.017$      & $0.069   \pm 0.016$      &  $0.068   \pm 0.015$      \\[+1mm]
  $\ln(10^{10} A_s)$       & $3.060 \pm 0.030$      & $3.066   \pm 0.029$      &  $3.063   \pm 0.027$      \\[+1mm]
  $n_s$                    & $0.9715 \pm 0.0057$    & $0.9702  \pm 0.0050$     &  $0.9696  \pm 0.0051$     \\[+1mm]
  $w$                      & $-1.24 \pm 0.20$       & $-0.979  \pm 0.047$      &  $-0.994  \pm 0.033$     \\[+1mm]
 \hline \\[-2mm]
  $H_0$ [km s$^{-1}$ Mpc$^{-1}$] & $76.7 \pm 7.0$   & $67.7  \pm 1.1$          &  $68.06   \pm 0.77$       \\[+1mm]
  $\Omega_m$               & $0.244 \pm 0.046$      & $0.307 \pm 0.010$        &  $0.3034  \pm 0.0073$     \\[+1mm]
  $\sigma_8$               & $0.877 \pm 0.054$      & $0.806 \pm 0.013$        &  $0.810   \pm 0.011$      \\[+0mm]
\end{tabular}
\end{ruledtabular}
\label{tab:para_flat_lensing}
\end{table*}

\begin{figure*}
\centering
\mbox{\includegraphics[width=65mm]{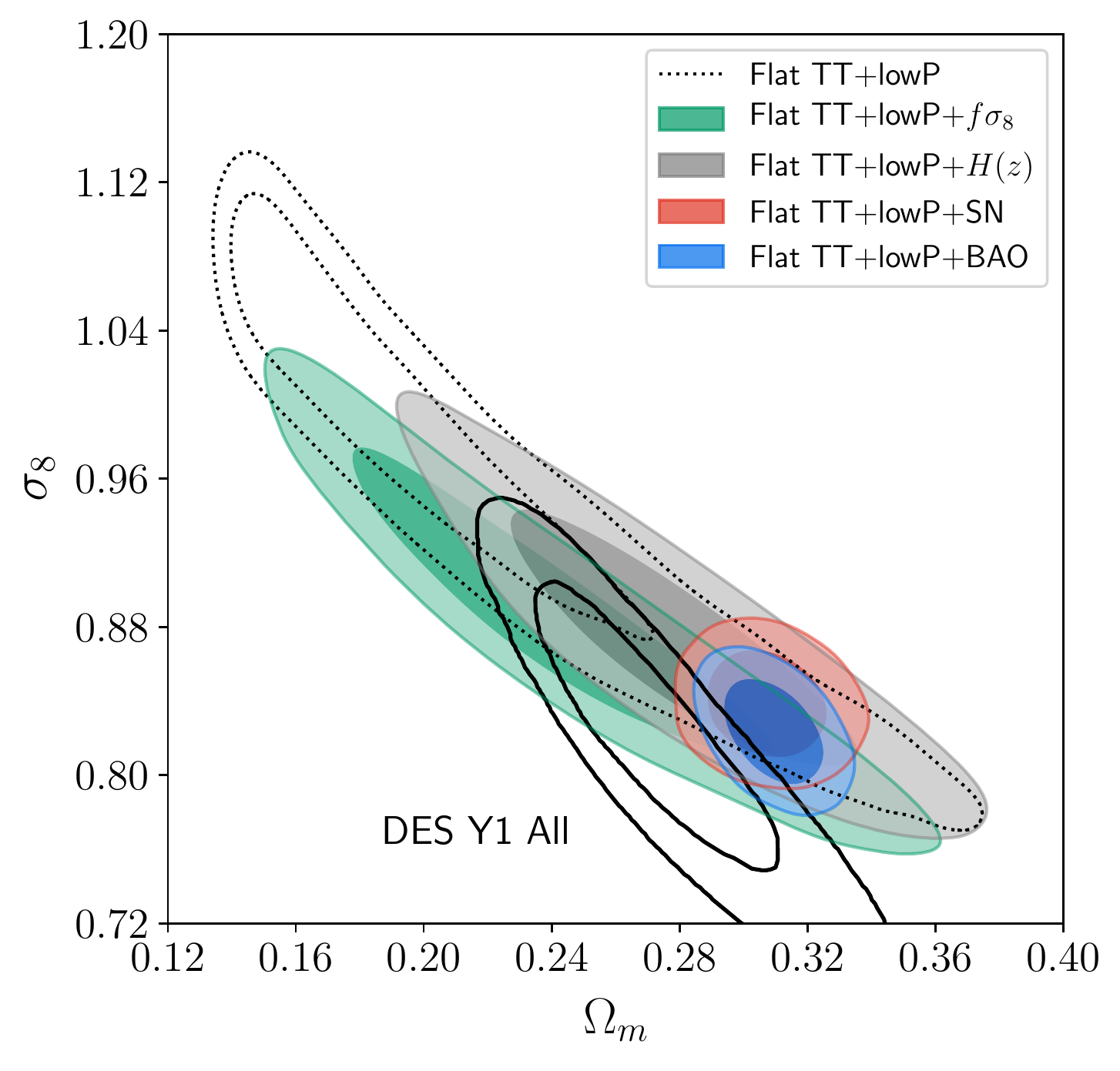}}
\mbox{\includegraphics[width=65mm]{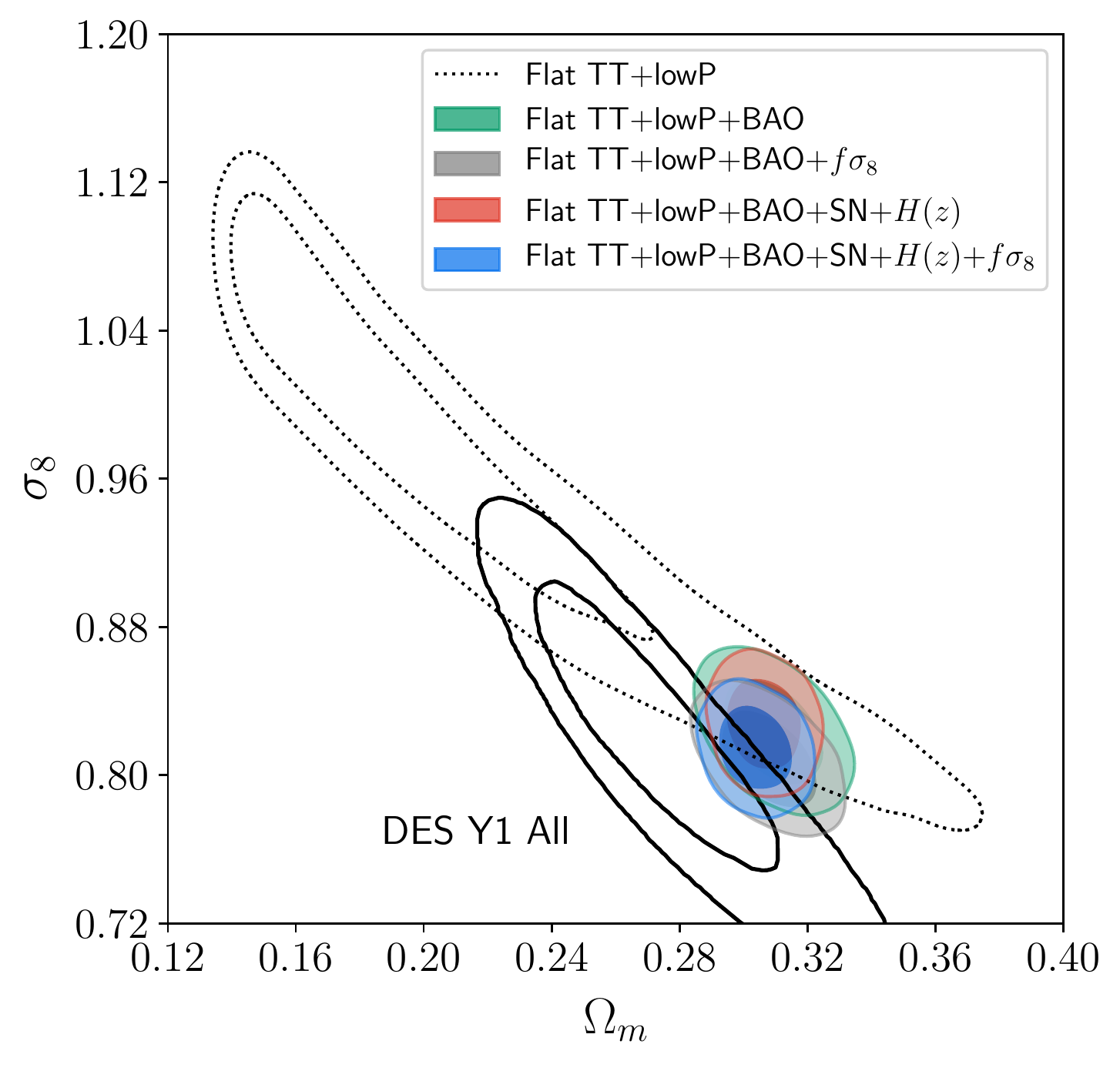}} \\
\mbox{\includegraphics[width=65mm]{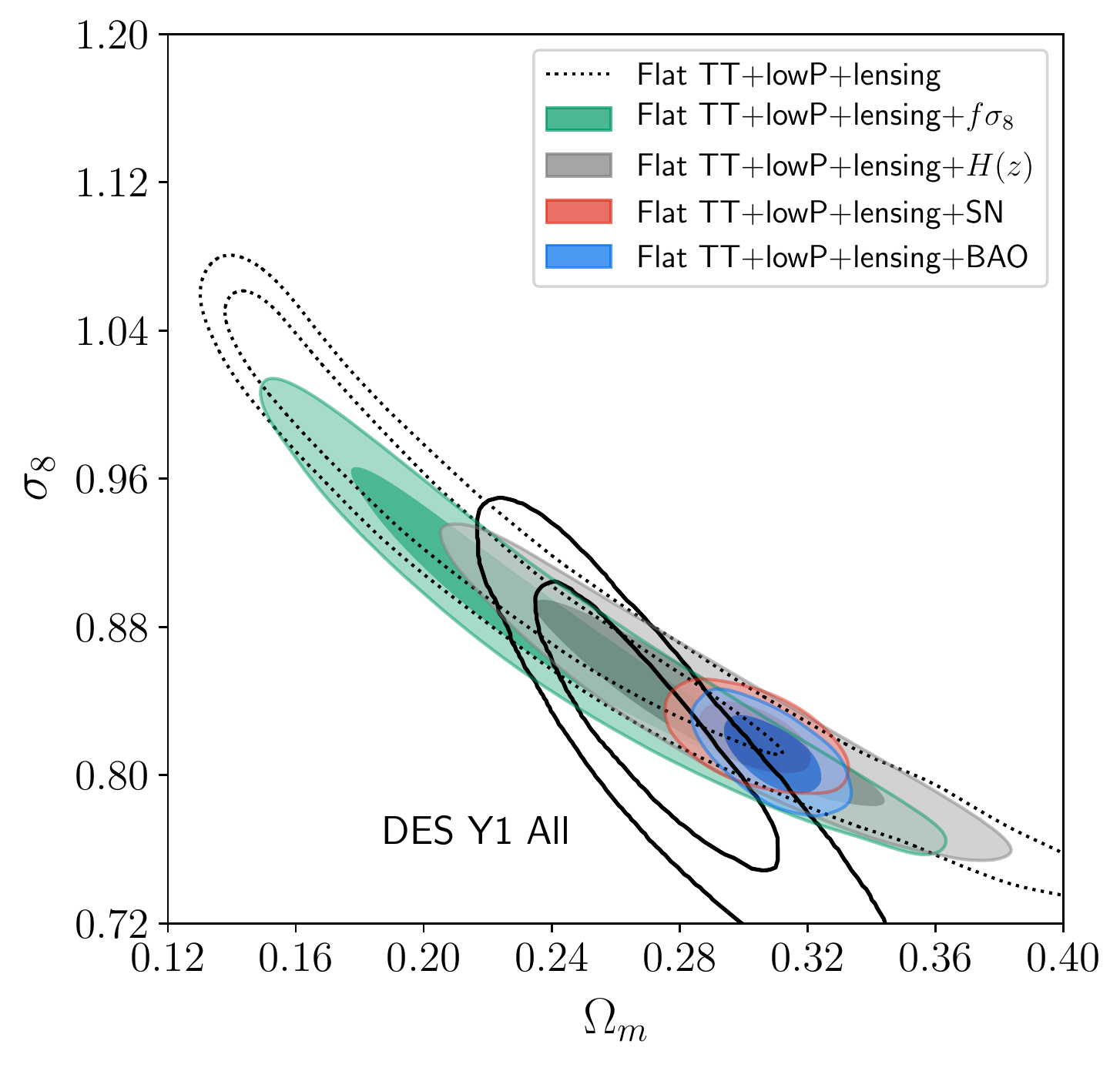}}
\mbox{\includegraphics[width=65mm]{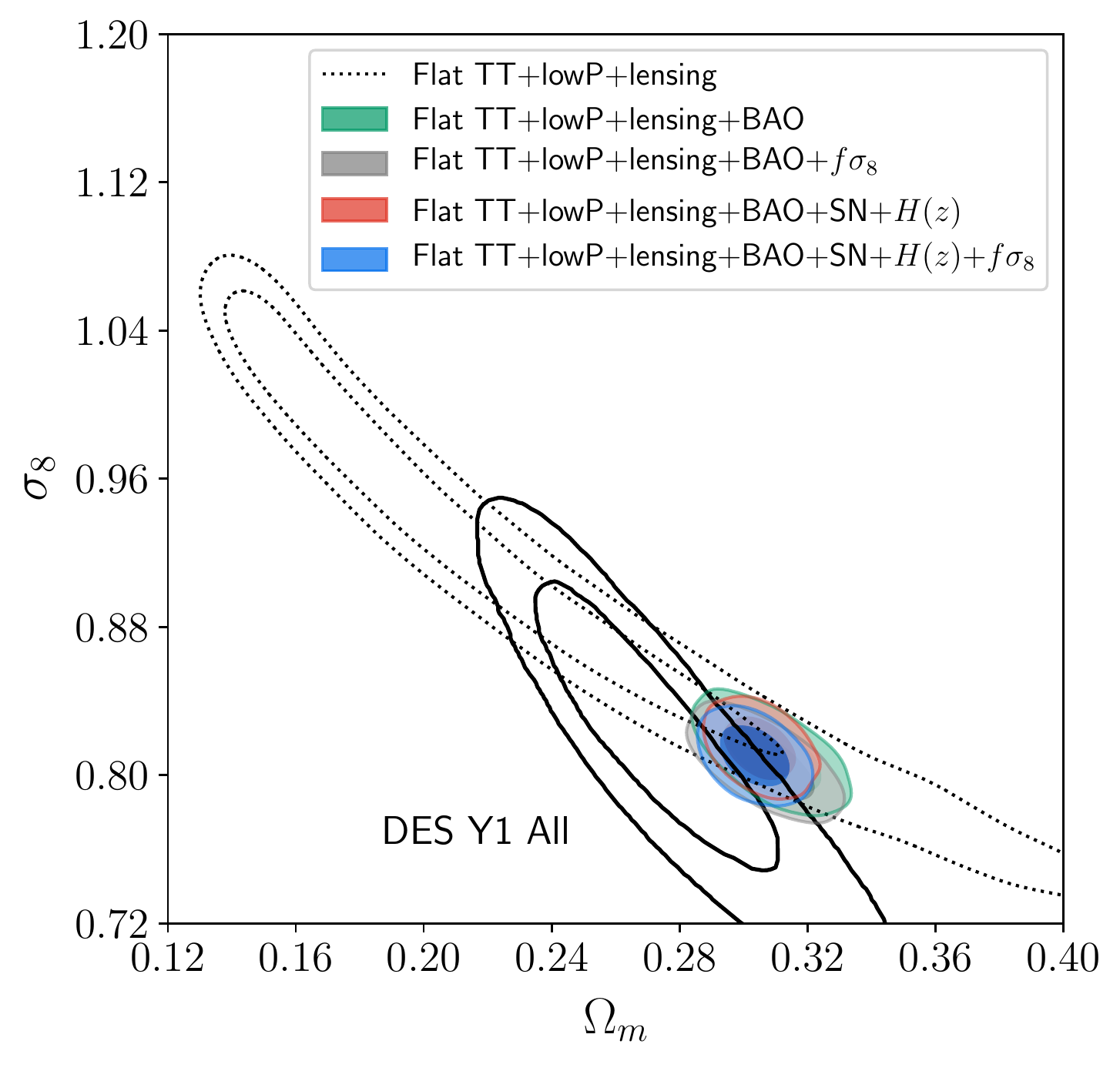}}
\caption{1$\sigma$ and 2$\sigma$ likelihood contours in the $\Omega_m$--$\sigma_8$ plane for
the tilted flat-XCDM parameterization constrained by Planck CMB TT + lowP (+lensing), SNIa, BAO, $H(z)$, and $f\sigma_8$ data.
In each panel the $\Lambda\textrm{CDM}$ model 1$\sigma$ and 2$\sigma$ 
constraint contours obtained from the first-year Dark Energy Survey (DES Y1 All) \citep{DESCollaboration2018}
are shown as thick solid curves for comparison.
}
\label{fig:omm_sig8_flat}
\end{figure*}

\begin{figure*}
\centering
\mbox{\includegraphics[width=65mm]{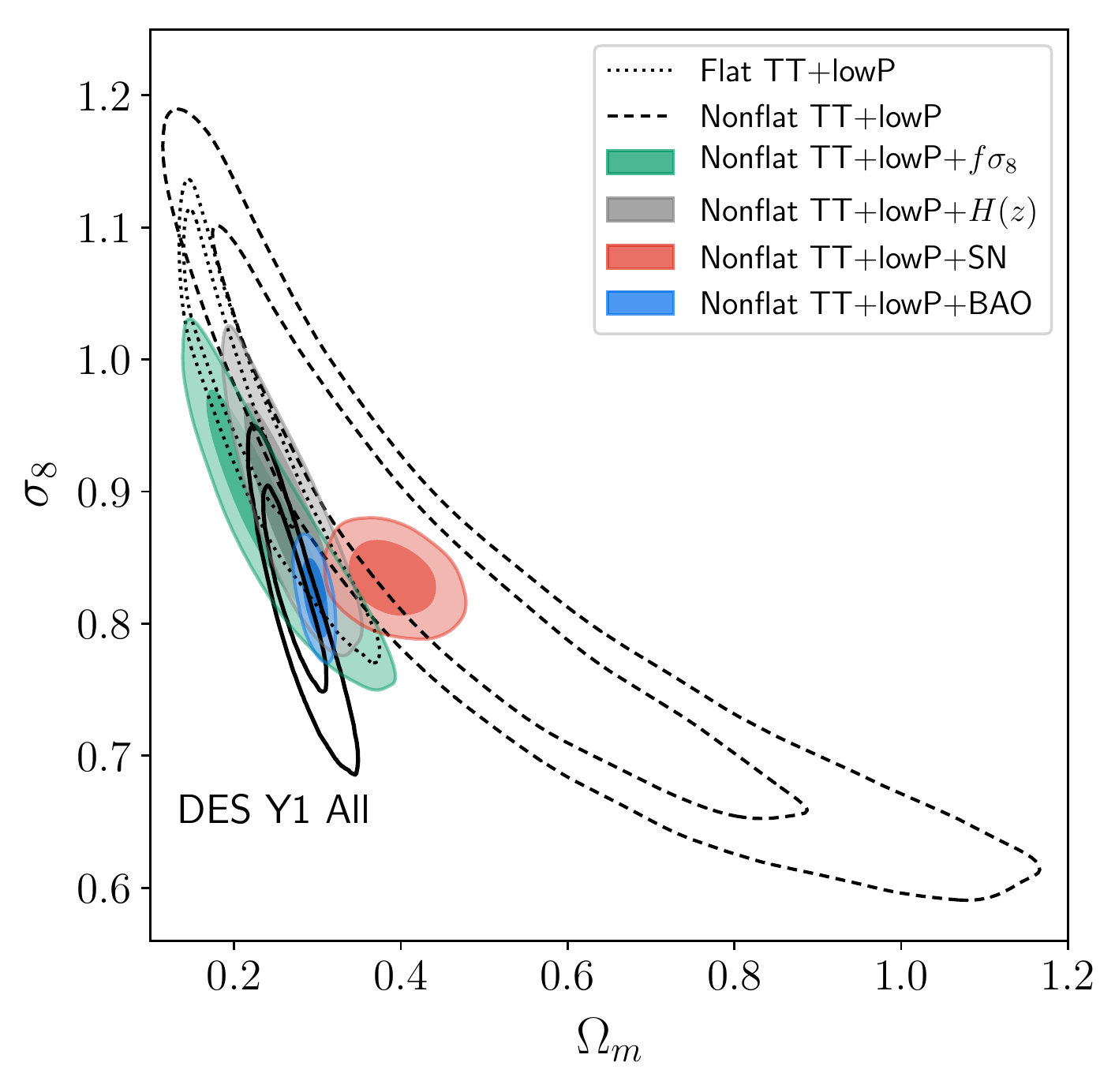}}
\mbox{\includegraphics[width=65mm]{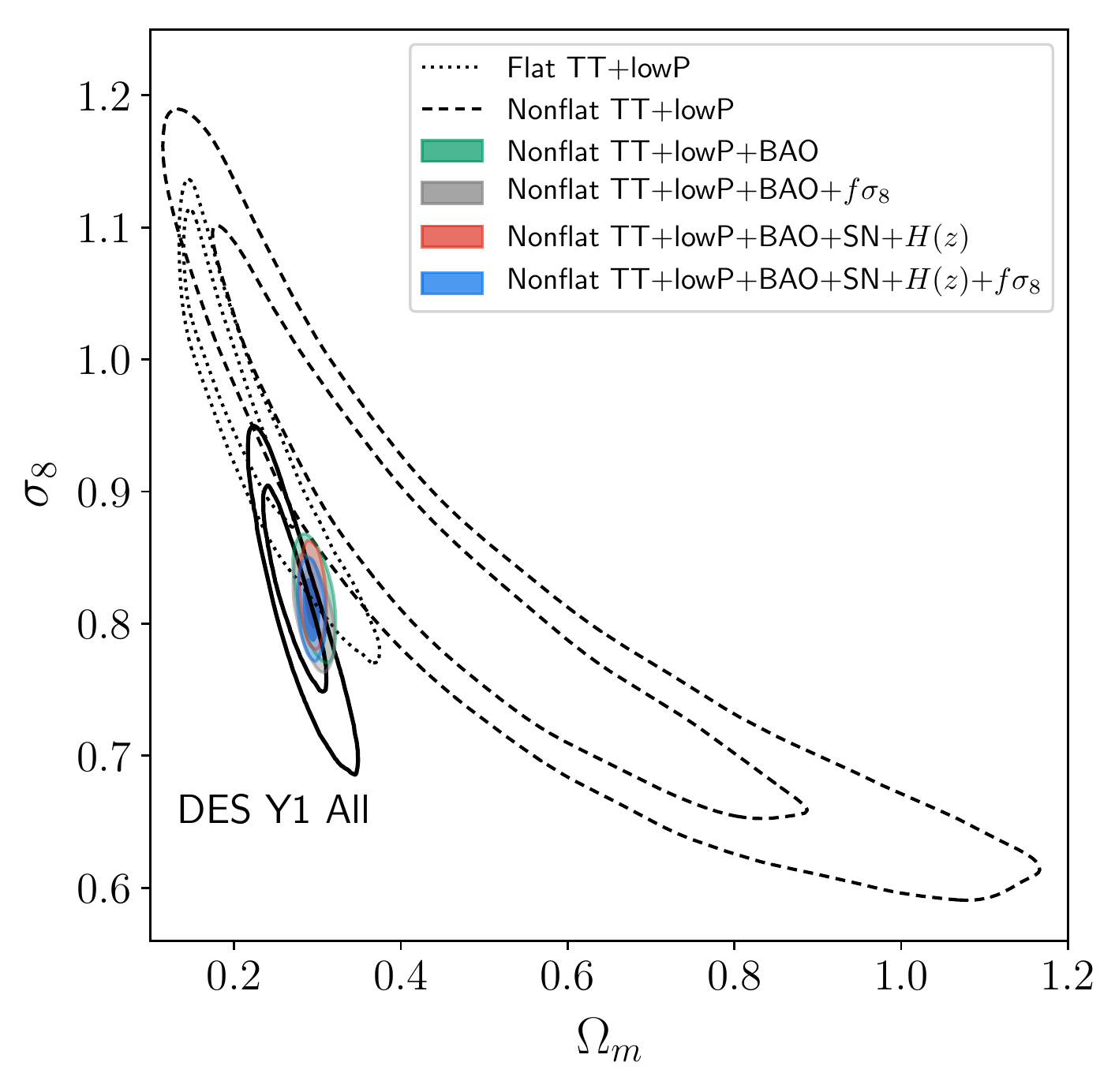}} \\
\mbox{\includegraphics[width=65mm]{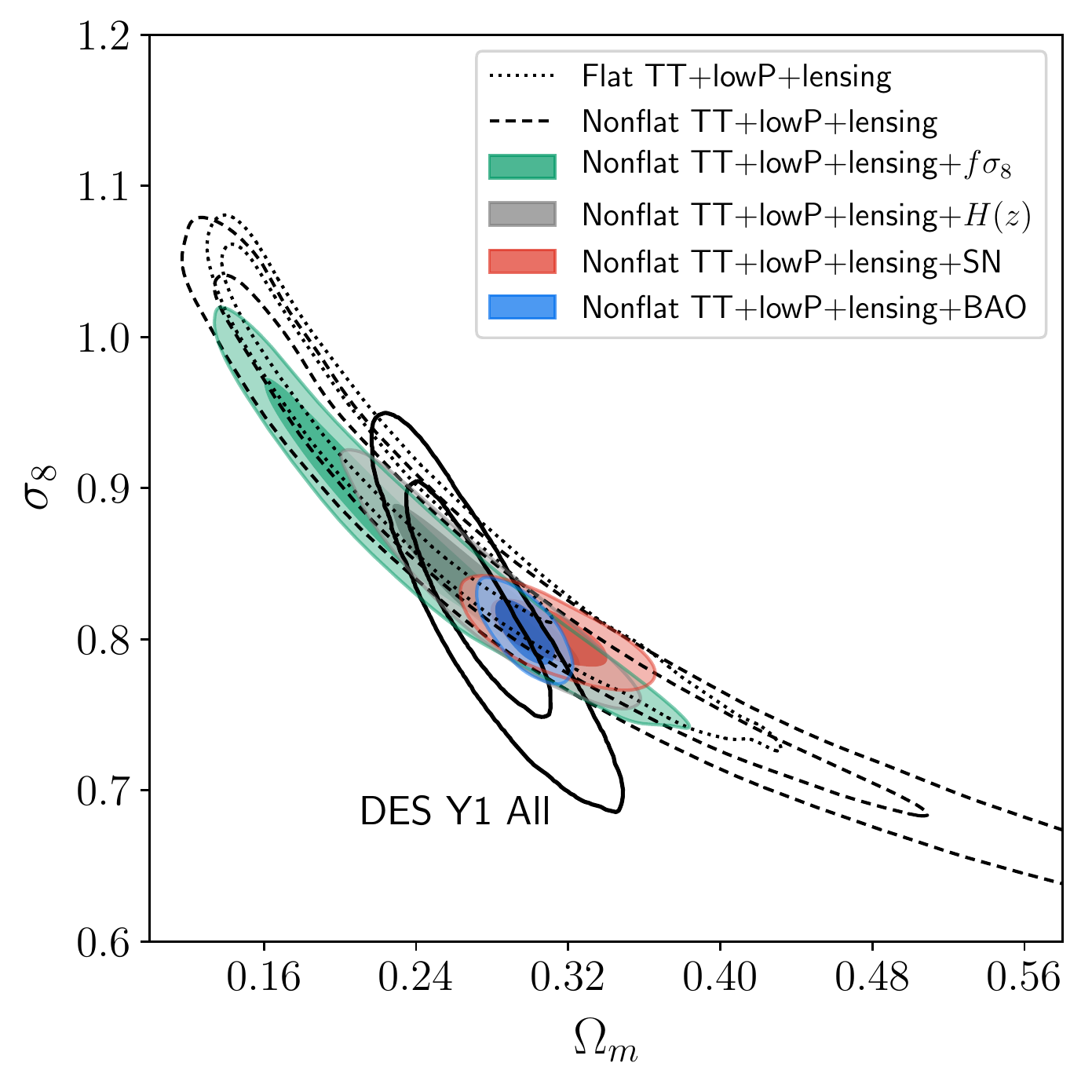}}
\mbox{\includegraphics[width=65mm]{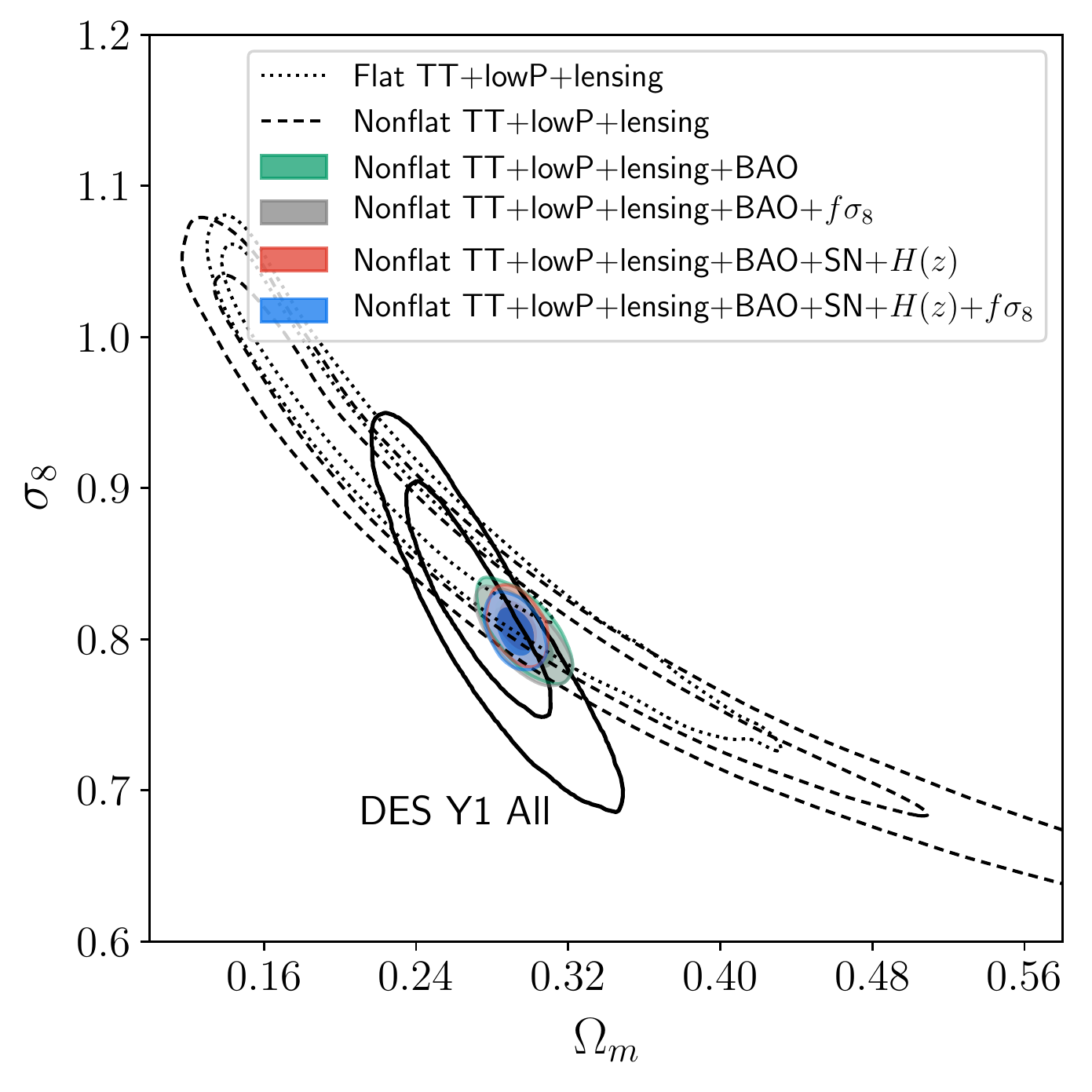}} 
\caption{Same as Fig.\ \ref{fig:omm_sig8_flat} but for the untilted nonflat
XCDM parameterization.
}
\label{fig:omm_sig8_nonflat}
\end{figure*}

For the full data combination (including CMB lensing data), $\sigma_8$'s 
measured using the two XCDM parameterizations, Tables 
\ref{tab:para_flat_lensing} and \ref{tab:para_nonflat_lensing}, 
agree to 0.32$\sigma$ (of the quadrature 
sum of the two error bars). Figures \ref{fig:omm_sig8_flat} and 
\ref{fig:omm_sig8_nonflat} show the marginalized two-dimensional likelihood 
distribution contours in the $\Omega_m$--$\sigma_8$ plane for the 
tilted flat and untilted nonflat XCDM parameterizations constrained using the 
CMB and non-CMB data.
For comparison we also plot the $\Lambda\textrm{CDM}$ constraints
obtained from a joint analysis of galaxy clustering and weak gravitational
lensing first year data of the Dark Energy Survey (DES Y1 All)
\citep{DESCollaboration2018}, whose 1$\sigma$ confidence ranges are 
$\Omega_m=0.264_{-0.019}^{+0.032}$ and $\sigma_8=0.807_{-0.041}^{+0.062}$.
The marginalized likelihood distribution contours in the 
$\Omega_m$--$\sigma_8$ plane determined by adding each non-CMB measurement 
set to the Planck 2015 CMB observations are consistent with each other, except 
for the nonflat XCDM parameterization where the TT + lowP + SN contours almost 
do not overlap 
with contours derived using any of the other three non-CMB data sets with the 
TT + lowP data (Fig.\ \ref{fig:omm_sig8_nonflat} top left panel). As expected, 
the BAO data provide the most restrictive constraints among the four 
non-CMB data sets.

While the $\sigma_8$ constraints from the tilted flat and untilted nonflat 
XCDM analyses (allowing for and ignoring CMB lensing data) are similar to the 
DES Y1 All result, the $\Omega_m$ constraints here favor a larger value by 
about $1.2\sigma$ (of the quadrature sum of the two error bars) for the 
flat-XCDM case for the full data combination. We emphasize that the best-fit 
point for the nonflat XCDM parameterization constrained by using the Planck 
CMB measurements (including lensing) combined with all non-CMB observations 
enters well into the 1$\sigma$ region of the DES Y1 All constraint contour 
(Fig.\ \ref{fig:omm_sig8_nonflat} lower right panel), unlike the tilted flat 
XCDM parameterization case (Fig.\ \ref{fig:omm_sig8_flat} 
lower right panel).


\begin{table*}
\caption{Untilted nonflat $\textrm{XCDM}$ model parameters constrained with Planck TT + lowP, SN, BAO, $H(z)$, and $f\sigma_8$ data (mean and 68.3\% confidence limits).}
\begin{ruledtabular}
\begin{tabular}{lccc}
  Parameter               &  TT+lowP                  &  TT+lowP+SN             &   TT+lowP+BAO    \\[+0mm]
 \hline \\[-2mm]
  $\Omega_b h^2$          &  $0.02335 \pm 0.00022$    &  $0.02328 \pm 0.00021$  &   $0.02305 \pm 0.00021$     \\[+1mm]
  $\Omega_c h^2$          &  $0.1093 \pm 0.0010$      &  $0.1093  \pm 0.0011$   &   $0.1097 \pm 0.0011$       \\[+1mm]
  $100\theta_\textrm{MC}$ &  $1.04240 \pm 0.00042$    &  $1.04237 \pm 0.00043$  &   $1.04222 \pm 0.00042$     \\[+1mm]
  $\tau$                  &  $0.087 \pm 0.029$        &  $0.107   \pm 0.022$    &   $0.134 \pm 0.017$       \\[+1mm]
  $\ln(10^{10} A_s)$      &  $3.083 \pm 0.058$        &  $3.124   \pm 0.044$    &   $3.178 \pm 0.034$        \\[+1mm]
  $\Omega_k$              &  $-0.084 \pm 0.052$       &  $-0.045  \pm 0.013$    &   $-0.0074 \pm 0.0024$     \\[+1mm]
  $w$                     &  $-1.45 \pm 0.75$         &  $-1.23   \pm 0.11$     &   $-0.959 \pm 0.056$     \\[+1mm]
 \hline \\[-2mm]
  $H_0$ [km s$^{-1}$ Mpc$^{-1}$] & $55 \pm 14$        &  $58.6 \pm 2.7$         &   $67.0 \pm 1.2$        \\[+1mm]
  $\Omega_m$              &  $0.52 \pm 0.24$          &  $0.390 \pm 0.035$      &   $0.297 \pm 0.010$     \\[+1mm]
  $\sigma_8$              &  $0.82 \pm 0.13$          &  $0.834 \pm 0.019$      &   $0.820 \pm 0.020$       \\[+1mm]
    \hline \hline \\[-2mm]
  Parameter               &  TT+lowP+$H(z)$           &  TT+lowP+SN+BAO         &  TT+lowP+SN+BAO+$H(z)$      \\[+0mm]
 \hline \\[-2mm]
  $\Omega_b h^2$          &  $0.02315 \pm 0.00020$    &  $0.02305 \pm 0.00020$  &  $0.02306  \pm 0.00020$       \\[+1mm]
  $\Omega_c h^2$          &  $0.1097 \pm 0.0011$      &  $0.1096  \pm 0.0011$   &  $0.1097   \pm 0.0010$      \\[+1mm]
  $100\theta_\textrm{MC}$ &  $1.04227 \pm 0.00042$    &  $1.04219 \pm 0.00041$  &  $1.04221  \pm 0.00041$       \\[+1mm]
  $\tau$                  &  $0.131 \pm 0.018$        &  $0.134   \pm 0.017$    &  $0.133    \pm 0.017$         \\[+1mm]
  $\ln(10^{10} A_s)$      &  $3.171 \pm 0.036$        &  $3.178   \pm 0.034$    &  $3.175    \pm 0.034$         \\[+1mm]
  $\Omega_k$              &  $-0.0122 \pm 0.0044$     &  $-0.0079 \pm 0.0021$   &  $-0.0074  \pm 0.0020$        \\[+1mm]
  $w$                     &  $-1.22 \pm 0.19$         &  $-0.974  \pm 0.0033$   &  $-0.968   \pm 0.033$        \\[+1mm]
 \hline \\[-2mm]
  $H_0$ [km s$^{-1}$ Mpc$^{-1}$]  & $71.6 \pm 4.7$    &  $67.26 \pm 0.80$       &  $67.34 \pm 0.74$             \\[+1mm]
  $\Omega_m$              &  $0.264 \pm 0.034$        &  $0.2949 \pm 0.0072$    &  $0.2944 \pm 0.0066$            \\[+1mm]
  $\sigma_8$              &  $0.888 \pm 0.050$        &  $0.824 \pm 0.017$      &  $0.822 \pm 0.017$           \\[+1mm]
   \hline \hline \\[-2mm]
     Parameter            &  TT+lowP+$f\sigma_8$      & TT+lowP+BAO+$f\sigma_8$ &  TT+lowP+SN+BAO+$H(z)$+$f\sigma_8$      \\[+0mm]
 \hline \\[-2mm]
  $\Omega_b h^2$          &  $0.02311 \pm 0.00020$    &  $0.02307 \pm 0.00019$  &  $0.02307  \pm 0.00020$       \\[+1mm]
  $\Omega_c h^2$          &  $0.1090  \pm 0.0010$     &  $0.1092  \pm 0.0011$   &  $0.1092   \pm 0.0010$      \\[+1mm]
  $100\theta_\textrm{MC}$ &  $1.04226 \pm 0.00041$    &  $1.04224 \pm 0.00041$  &  $1.04224  \pm 0.00042$       \\[+1mm]
  $\tau$                  &  $0.119 \pm 0.019$        &  $0.126   \pm 0.017$    &  $0.125    \pm 0.016$         \\[+1mm]
  $\ln(10^{10} A_s)$      &  $3.146 \pm 0.038$        &  $3.160   \pm 0.034$    &  $3.159    \pm 0.033$         \\[+1mm]
  $\Omega_k$              &  $-0.0089 \pm 0.0077$     &  $-0.0070 \pm 0.0024$   &  $-0.0071  \pm 0.0020$        \\[+1mm]
  $w$                     &  $-1.22 \pm 0.18$         &  $-0.951  \pm 0.054$    &  $-0.961   \pm 0.033$         \\[+1mm]
 \hline \\[-2mm]
  $H_0$ [km s$^{-1}$ Mpc$^{-1}$]  & $74.9 \pm 8.1$    &  $67.1  \pm 1.1$        &  $67.41 \pm 0.77$             \\[+1mm]
  $\Omega_m$              &  $0.245 \pm 0.054$        &  $0.295 \pm 0.010$      &  $0.2926 \pm 0.0068$            \\[+1mm]
  $\sigma_8$              &  $0.880 \pm 0.058$        &  $0.808 \pm 0.018$      &  $0.811 \pm 0.016$                \\[+0mm]
\end{tabular}
\end{ruledtabular}
\label{tab:para_nonflat}
\end{table*}

Table \ref{tab:chi2_lcdm} lists $\chi^2$
values for the best-fit tilted flat and untilted nonflat 
$\Lambda\textrm{CDM}$ models.
This is an updated version of Table 9 of \citet{ParkRatra2018a}, for the 
updated data sets we use here. Table \ref{tab:chi2} lists the corresponding 
quantities for the tilted flat and the untilted nonflat XCDM parameterizations.
The best-fit position in parameter space is determined from Powell's 
minimization method that
is an efficient algorithm to find the location of the minimum $\chi^2$. We use the COSMOMC program
(with an option {\tt action=2}) to implement this method.\footnote{Our $\chi^2$ values 
presented here for the tilted flat-$\textrm{XCDM}$ model constrained with TT + lowP data
are similar to the Planck results ($\chi_\textrm{PlikTT}^2=761.9$,
$\chi_\textrm{lowTEB}^2=10495.14$, $\chi_\textrm{prior}^2=1.86$
with total $\chi^2=11258.91$; \citealt{PlanckParameterTables2015}).}
In these Tables we list the individual $\chi^2$ contribution of each data
set used to constrain model parameters. 
The total $\chi^2$ is the sum of those of the high-$\ell$ CMB TT 
likelihood ($\chi_{\textrm{PlikTT}}^2$),
the low-$\ell$ CMB power spectra of temperature and polarization
($\chi_{\textrm{lowTEB}}^2$), lensing ($\chi_{\textrm{lensing}}^2$),
SN ($\chi_{\textrm{SN}}^2$), $H(z)$ ($\chi_{H(z)}^2$), BAO ($\chi_{\textrm{BAO}}^2$),
$f\sigma_8$ data ($\chi_{f\sigma_8}^2$), as well as
the contribution from the foreground nuisance parameters
($\chi_{\textrm{prior}}^2$).
Due to the nonconventional normalization of the Planck CMB anisotropy data 
likelihoods, the number of Planck 2015 CMB degrees of freedom is ambiguous. 
Given that the number of degrees of freedom of the Planck 2015 CMB data is 
unavailable and that the absolute value of $\chi^2$ is arbitrary, only 
$\chi^2$ differences between two models are meaningful 
for the Planck CMB data. In Table \ref{tab:chi2_lcdm}, for the untilted nonflat 
$\Lambda\textrm{CDM}$ model, we list $\Delta\chi^2$, the excess $\chi^2$ 
relative to the value of the tilted flat-$\Lambda\textrm{CDM}$ model 
constrained 
using the same combination of measurements. For the non-CMB observations, the 
numbers of degrees of freedom are 1042, 31, 15, 10 for the SN, $H(z)$, BAO, 
$f\sigma_8$ data sets, respectively, a total of 1098 degrees of
freedom. The reduced $\chi^2$'s for each of the non-CMB measurement sets are 
$\chi^2 / \nu \lesssim 1$. There are 189 points in the Planck 2015 (binned
angular power spectrum) TT + lowP data and 197 when the CMB lensing
data are included. 

Conclusions about the qualitative relative goodness of fit of the 
tilted flat and nonflat $\Lambda\textrm{CDM}$ models drawn from the 
updated data here are not very different from those found earlier 
\citep{ParkRatra2018a} from the original data. For the nonflat 
$\Lambda$CDM case relative to the flat-$\Lambda$CDM model,
we have $\Delta\chi^2=21$ for  TT + lowP + lensing and the full non-CMB 
compilation (last column in the last row of Table \ref{tab:chi2_lcdm}).
As discussed above and in \citet{Oobaetal2018a, Oobaetal2018b, Oobaetal2018c}
and \citet{ParkRatra2018a, ParkRatra2018b}, it is not clear how to convert 
this into a 
quantitative relative probability as the two six parameter cases are not 
nested (and the number of degrees of freedom of the Planck measurements
is unavailable). It is clear however that the nonflat $\Lambda$CDM model 
does a worse job in fitting the higher-$\ell$ $C_\ell$'s than it does in 
fitting the lower-$\ell$ ones. We note that there has been discussion 
about systematic differences between constraints determined 
using the higher-$\ell$ and the lower-$\ell$ Planck 2015 CMB data \citep{Addisonetal2016,PlanckCollaboration2017}. In addition, in the context of the 
flat-$\Lambda$CDM model, there appear to be inconsistencies between the 
higher-$\ell$ Planck 2015 CMB anisotropy data and the South Pole Telescope 
CMB anisotropy data \citep{Ayloretal2017}. It is possible that, if real, when
these differences are resolved this could result in a reduction of the 
$\Delta\chi^2$'s found here.


\begin{table*}
\caption{Untilted nonflat $\textrm{XCDM}$ model parameters constrained with Planck TT + lowP + lensing, SN, BAO, $H(z)$, and $f\sigma_8$ data (mean and 68.3\% confidence limits).}
\begin{ruledtabular}
\begin{tabular}{lccc}
  Parameter               &  TT+lowP+lensing          & TT+lowP+lensing+SN      &   TT+lowP+lensing+BAO      \\[+0mm]
 \hline \\[-2mm]
  $\Omega_b h^2$          &  $0.02305 \pm 0.00020$    &  $0.02305 \pm 0.00019$  &   $0.02302 \pm 0.00020$       \\[+1mm]
  $\Omega_c h^2$          &  $0.1091 \pm 0.0011$      &  $0.1091 \pm 0.0011$    &   $0.1094 \pm 0.0011$         \\[+1mm]
  $100\theta_\textrm{MC}$ &  $1.04235 \pm 0.00043$    &  $1.04234 \pm 0.00041$  &   $1.04226 \pm 0.00041$       \\[+1mm]
  $\tau$                  &  $0.100 \pm 0.022$        &  $0.101 \pm 0.021$      &   $0.123 \pm 0.014$        \\[+1mm]
  $\ln(10^{10} A_s)$      &  $3.106 \pm 0.044$        &  $3.109 \pm 0.042$      &   $3.154 \pm 0.027$           \\[+1mm]
  $\Omega_k$              &  $-0.019 \pm 0.017$       &  $-0.0153 \pm 0.0075$   &   $-0.0070 \pm 0.0025$        \\[+1mm]
  $w$                     &  $-1.12 \pm 0.39$         &  $-1.019  \pm 0.053$    &   $-0.946   \pm 0.056$         \\[+1mm]
 \hline \\[-2mm]
  $H_0$ [km s$^{-1}$ Mpc$^{-1}$] &  $69 \pm 14$       &  $65.5 \pm 2.3$         &   $66.9 \pm 1.2$            \\[+1mm]
  $\Omega_m$              &  $0.31 \pm 0.13$          &  $0.310 \pm 0.021$      &   $0.297 \pm 0.010$          \\[+1mm]
  $\sigma_8$              &  $0.83 \pm 0.11$          &  $0.803 \pm 0.015$      &   $0.805  \pm 0.014$         \\[+1mm]
  \hline\hline \\[-2mm]
  Parameter               &  TT+lowP+lensing+$H(z)$   &  TT+lowP+lensing+SN+BAO &  TT+lowP+lensing+SN+BAO+$H(z)$      \\[+0mm]
 \hline \\[-2mm]
  $\Omega_b h^2$          &  $0.02304 \pm 0.00020$    &  $0.02303 \pm 0.00019$  &  $0.02303 \pm 0.00021$   \\[+1mm]
  $\Omega_c h^2$          &  $0.1094 \pm 0.0011$      &  $0.1094  \pm 0.0011$   &  $0.1095 \pm 0.0010$    \\[+1mm]
  $100\theta_\textrm{MC}$ &  $1.04232 \pm 0.00043$    &  $1.04228 \pm 0.00040$  &  $1.04228 \pm 0.00041$     \\[+1mm]
  $\tau$                  &  $0.114 \pm 0.017$        &  $0.120   \pm 0.012$    &  $0.121 \pm 0.012$       \\[+1mm]
  $\ln(10^{10} A_s)$      &  $3.137 \pm 0.034$        &  $3.148   \pm 0.024$    &  $3.150 \pm 0.024$       \\[+1mm]
  $\Omega_k$              &  $-0.0081 \pm 0.0042$     &  $-0.0075 \pm 0.0021$   &  $-0.0070 \pm 0.0020$      \\[+1mm]
  $w$                     &  $-1.06 \pm 0.14$         &  $-0.967  \pm 0.033$    &  $-0.961 \pm 0.033$      \\[+1mm]
 \hline \\[-2mm]
  $H_0$ [km s$^{-1}$ Mpc$^{-1}$]  &  $69.9 \pm 4.2$   &  $67.31 \pm 0.77$       &  $67.38 \pm 0.75$   \\[+1mm]
  $\Omega_m$              &  $0.275 \pm 0.033$        &  $0.2938 \pm 0.0069$    &  $0.2933 \pm 0.0067$      \\[+1mm]
  $\sigma_8$              &  $0.832  \pm 0.036$       &  $0.809 \pm 0.011$      &  $0.809 \pm 0.011$    \\[+1mm]
    \hline\hline \\[-2mm]
  Parameter               &  TT+lowP+lensing+$f\sigma_8$    &  TT+lowP+lensing+BAO+$f\sigma_8$   &  TT+lowP+lensing+SN+BAO+$H(z)$+$f\sigma_8$  \\[+0mm]
 \hline \\[-2mm]
  $\Omega_b h^2$          &  $0.02306 \pm 0.00020$    &  $0.02303 \pm 0.00020$  & $0.02305 \pm 0.00020$   \\[+1mm]
  $\Omega_c h^2$          &  $0.1090 \pm 0.0011$      &  $0.1093  \pm 0.0011$   & $0.1092 \pm 0.0010$      \\[+1mm]
  $100\theta_\textrm{MC}$ &  $1.04229 \pm 0.00041$    &  $1.04225 \pm 0.00043$  & $1.04227 \pm 0.00042$  \\[+1mm]
  $\tau$                  &  $0.114 \pm 0.019$        &  $0.120   \pm 0.013$    & $0.119 \pm 0.012$        \\[+1mm]
  $\ln(10^{10} A_s)$      &  $3.136 \pm 0.038$        &  $3.148   \pm 0.026$    & $3.146 \pm 0.024$    \\[+1mm]
  $\Omega_k$              &  $-0.0056 \pm 0.0063$     &  $-0.0068 \pm 0.0024$   & $-0.0069 \pm 0.0020$  \\[+1mm]
  $w$                     &  $-1.19 \pm 0.18$         &  $-0.943  \pm 0.054$    & $-0.960 \pm 0.032$    \\[+1mm]
 \hline \\[-2mm]
  $H_0$ [km s$^{-1}$ Mpc$^{-1}$]  &  $76.4 \pm 8.3$   &  $67.0 \pm 1.1$         &  $67.45 \pm 0.75$     \\[+1mm]
  $\Omega_m$              &  $0.236 \pm 0.053$        &  $0.297 \pm 0.010$      &  $0.2923 \pm 0.0066$         \\[+1mm]
  $\sigma_8$              &  $0.873 \pm 0.059$        &  $0.801 \pm 0.014$      &  $0.805 \pm 0.011$  \\[+0mm]
\end{tabular}
\end{ruledtabular}
\label{tab:para_nonflat_lensing}
\end{table*}

Table \ref{tab:chi2} lists $\chi^2$ values for the 
best-fit tilted flat and untilted nonflat XCDM parameterizations. In the 
last column we list $\Delta\chi^2$, the excess $\chi^2$ of the seven parameter
XCDM case over the value of the corresponding six parameter 
$\Lambda\textrm{CDM}$ model constrained using the same combination of 
data sets.
These models are nested; the seven parameter tilted flat-XCDM (untilted 
nonflat XCDM) parameterization reduces to the six parameter tilted 
flat-$\Lambda$CDM (untilted nonflat $\Lambda$CDM) model when $w = -1$. In 
this case the ambiguity in the number of Planck 2015 degrees of freedom is
not an obstacle to converting the $\Delta\chi^2$ values to a relative 
goodness of fit. From $\sqrt{-\Delta \chi^2}$, for the full data
set (including CMB lensing), for one additional free parameter, we find 
that the  tilted flat-XCDM (untilted nonflat XCDM) parameterization
is a 0.28$\sigma$ (0.87$\sigma$) better fit to the data than is the tilted 
flat-$\Lambda$CDM (untilted nonflat $\Lambda$CDM) model. (We emphasize 
that nonflat $\Lambda$CDM does not fit the data as well as flat-$\Lambda$CDM,
although the difference in the goodness of fit cannot yet be precisely 
quantified.) These results are consistent with those of \citet{Oobaetal2018d} 
and \citet{Oobaetal2018b}. 

Of all these four models, the tilted flat-XCDM parameterization best fits the 
combined data, but at a lower level of significance than the 1.1$\sigma$ of 
\citet{Oobaetal2018d}, and not close to the 3 or 4$\sigma$ significance 
found in earlier approximate analyses \citep{Solaetal2017a, Solaetal2018,
Solaetal2017b, Solaetal2017c, Solaetal2017d, GomezValentSola2017, GomezValentSola2018}.
While the tilted flat-XCDM parameterization does not provide a significantly
better fit to the data, current data cannot rule out dynamical dark energy.

\begin{figure*}[htbp]
\centering
\mbox{\includegraphics[width=100mm,bb=15 230 570 750]{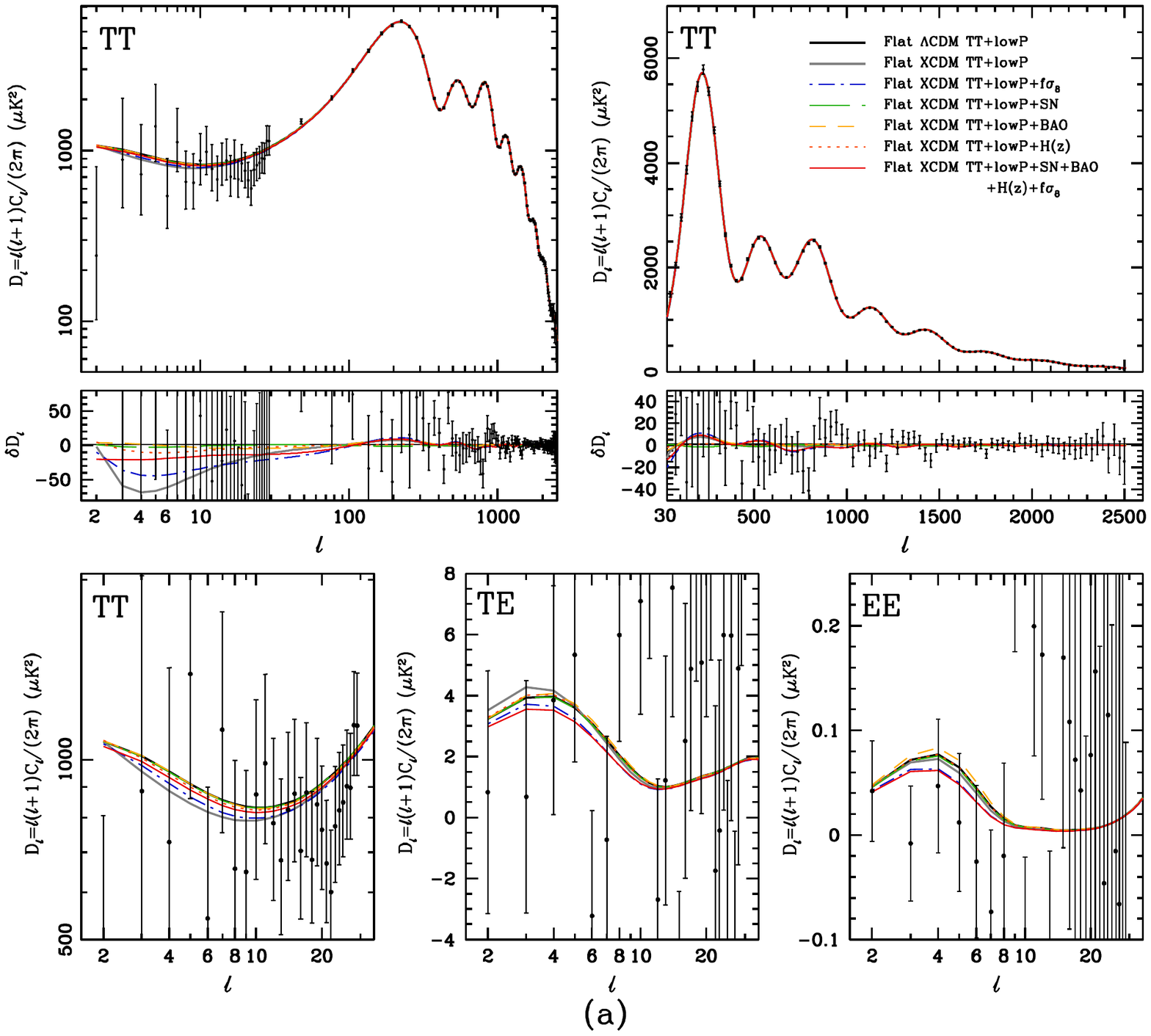}} \\
\mbox{\includegraphics[width=100mm,bb=15 210 570 750]{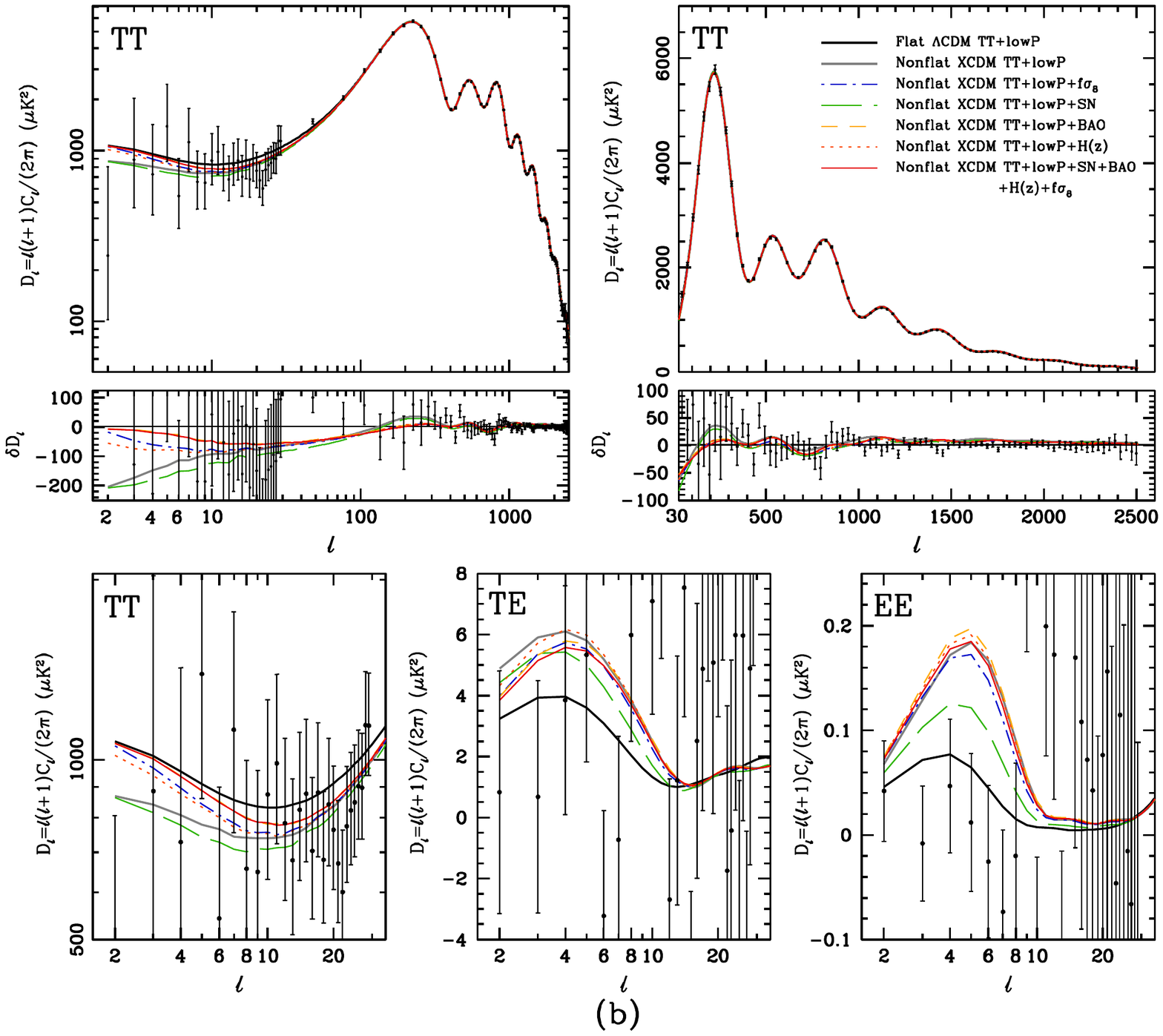}} 
\caption{Best-fit CMB anisotropy power spectra of (a) tilted flat (top five panels) and (b)
untilted nonflat $\textrm{XCDM}$ parameterizations (bottom five panels) constrained
by the Planck CMB TT + lowP data (excluding the lensing data) together with SN, BAO, $H(z)$, and $f\sigma_8$ data.
For comparison, the best-fit power spectra of the tilted flat-$\Lambda\textrm{CDM}$ model are shown as 
black curves.
The residual $\delta D_\ell$ of the TT power spectra are shown with respect to the
flat-$\Lambda\textrm{CDM}$ power spectrum that best fits the TT + lowP data. 
}
\label{fig:ps_cmb}
\end{figure*}

\begin{figure*}[htbp]
\centering
\mbox{\includegraphics[width=100mm,bb=15 230 570 750]{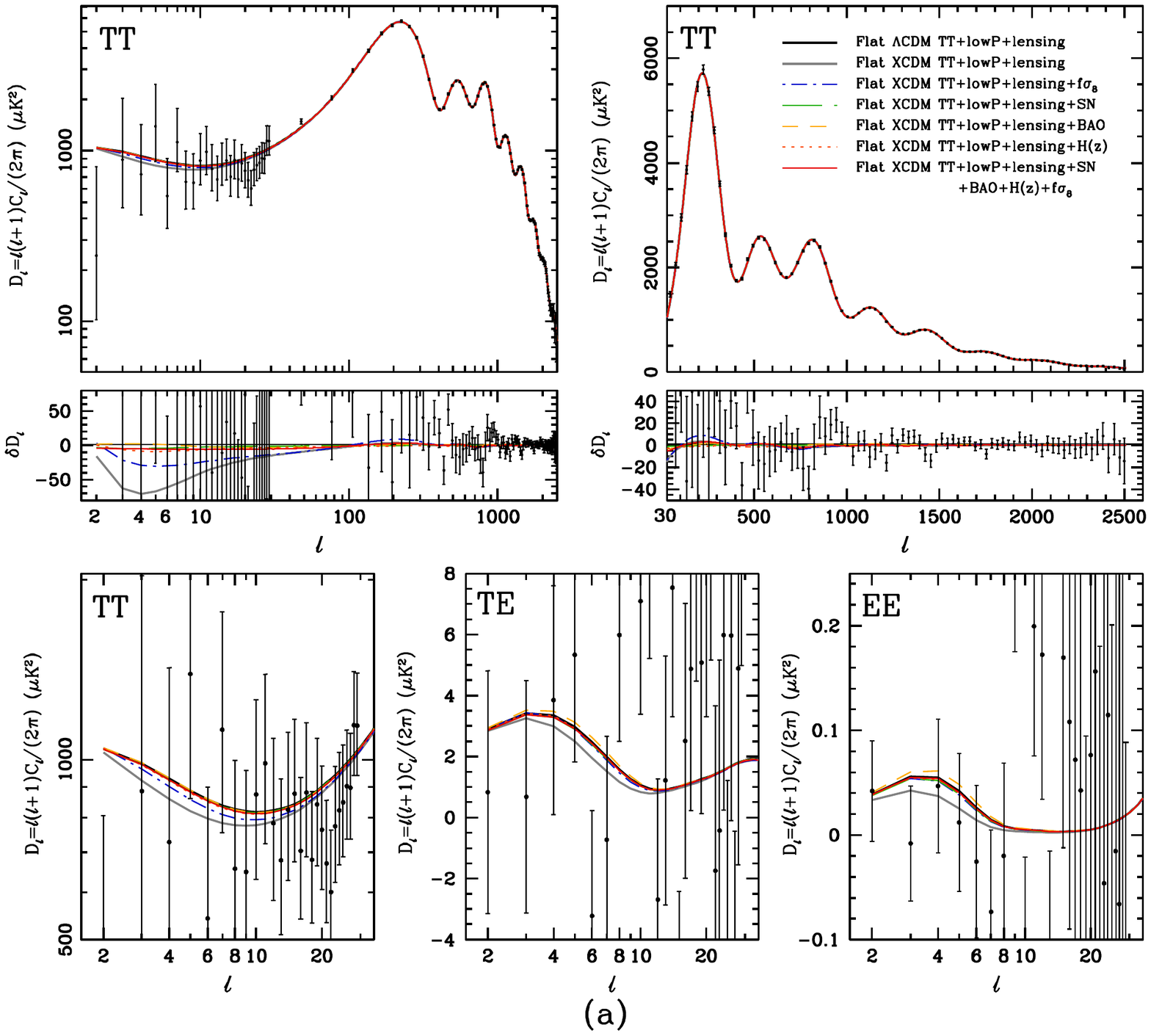}} \\
\mbox{\includegraphics[width=100mm,bb=15 210 570 750]{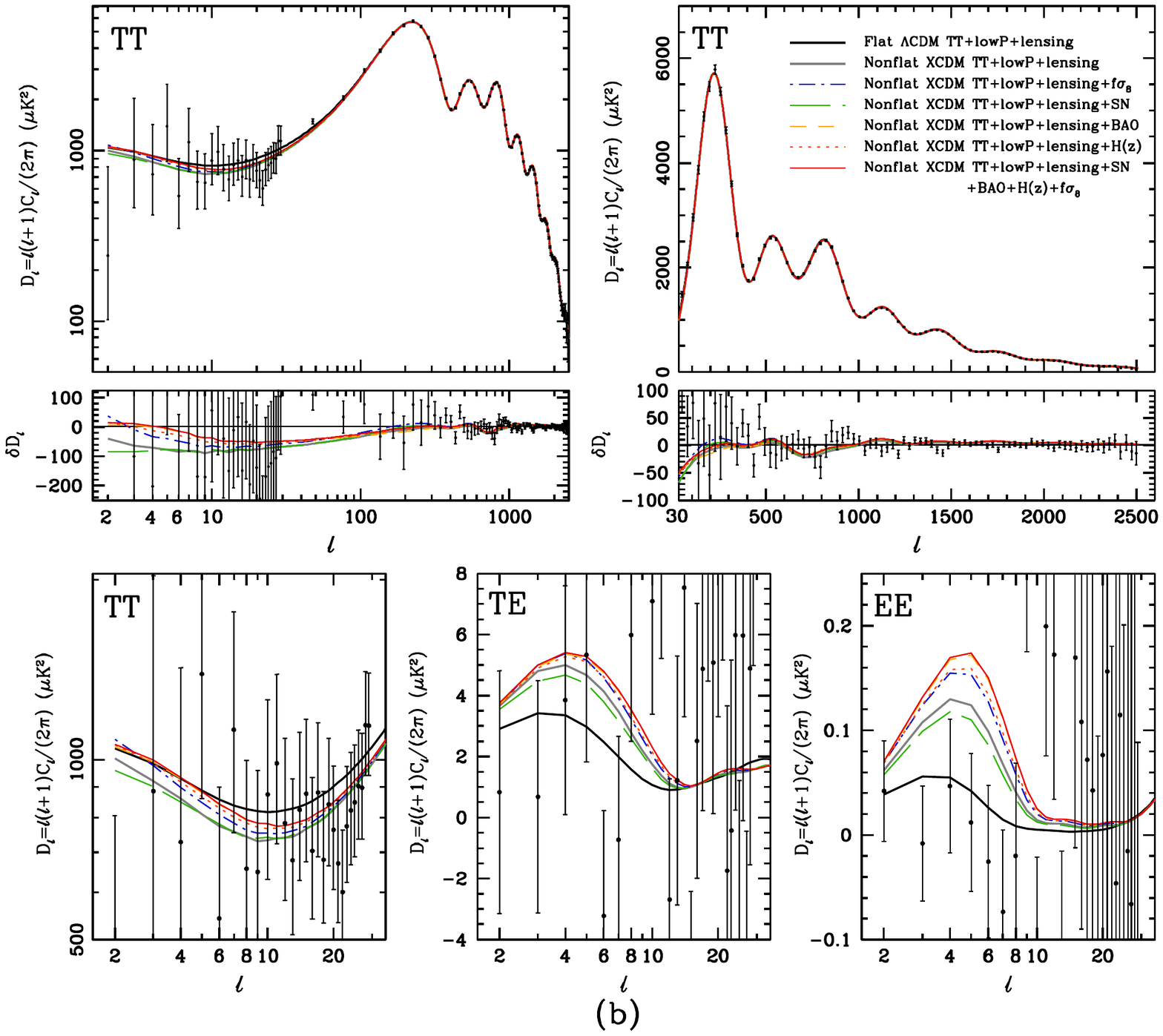}}
\caption{Same as Fig. \ref{fig:ps_cmb} but now including the CMB lensing data.
The residual $\delta D_\ell$ of the TT power spectra are shown with respect to the
flat-$\Lambda\textrm{CDM}$ power spectrum that best fits the TT + lowP + lensing data. 
}
\label{fig:ps_cmb_lensing}
\end{figure*}

Figures \ref{fig:ps_cmb} and \ref{fig:ps_cmb_lensing} show plots of the CMB 
high-$\ell$ TT, and the low-$\ell$ TT, TE, EE power spectra of the best-fit 
tilted flat and untilted nonflat XCDM dynamical dark energy inflation 
parameterizations, excluding and including the lensing data, respectively. 
The best-fit tilted flat-XCDM models favored by the Planck CMB and non-CMB data
agree well with the observed CMB power spectra at all $\ell$.
However, similar to the nonflat $\Lambda\textrm{CDM}$ case
studied in \citet{ParkRatra2018a}, the nonflat XCDM parameterization 
constrained by using 
the Planck 2015 CMB data and each non-CMB data set generally
has a poorer fit to the low-$\ell$ EE anisotropy power spectrum while it 
better fits the low-$\ell$ TT anisotropy power spectrum (see the bottom left 
panel of Figs.\ 
\ref{fig:ps_cmb} and \ref{fig:ps_cmb_lensing}). The shape of the best-fit 
$C_\ell$ power spectra of models relative to the Planck CMB data 
points are quite consistent with the $\chi^2$'s listed in Table \ref{tab:chi2}.
For example, the best-fit untilted nonflat XCDM parameterization
constrained by using the TT + lowP + lensing and full non-CMB data sets
has a low-$\ell$ EE power spectrum that deviates the most from the Planck
data and the corresponding value of $\chi^2_{\textrm{lowTEB}}$ is larger
than values from other non-CMB combinations.

\begin{figure*}[htbp]
\mbox{\includegraphics[width=85mm,bb=30 170 500 620]{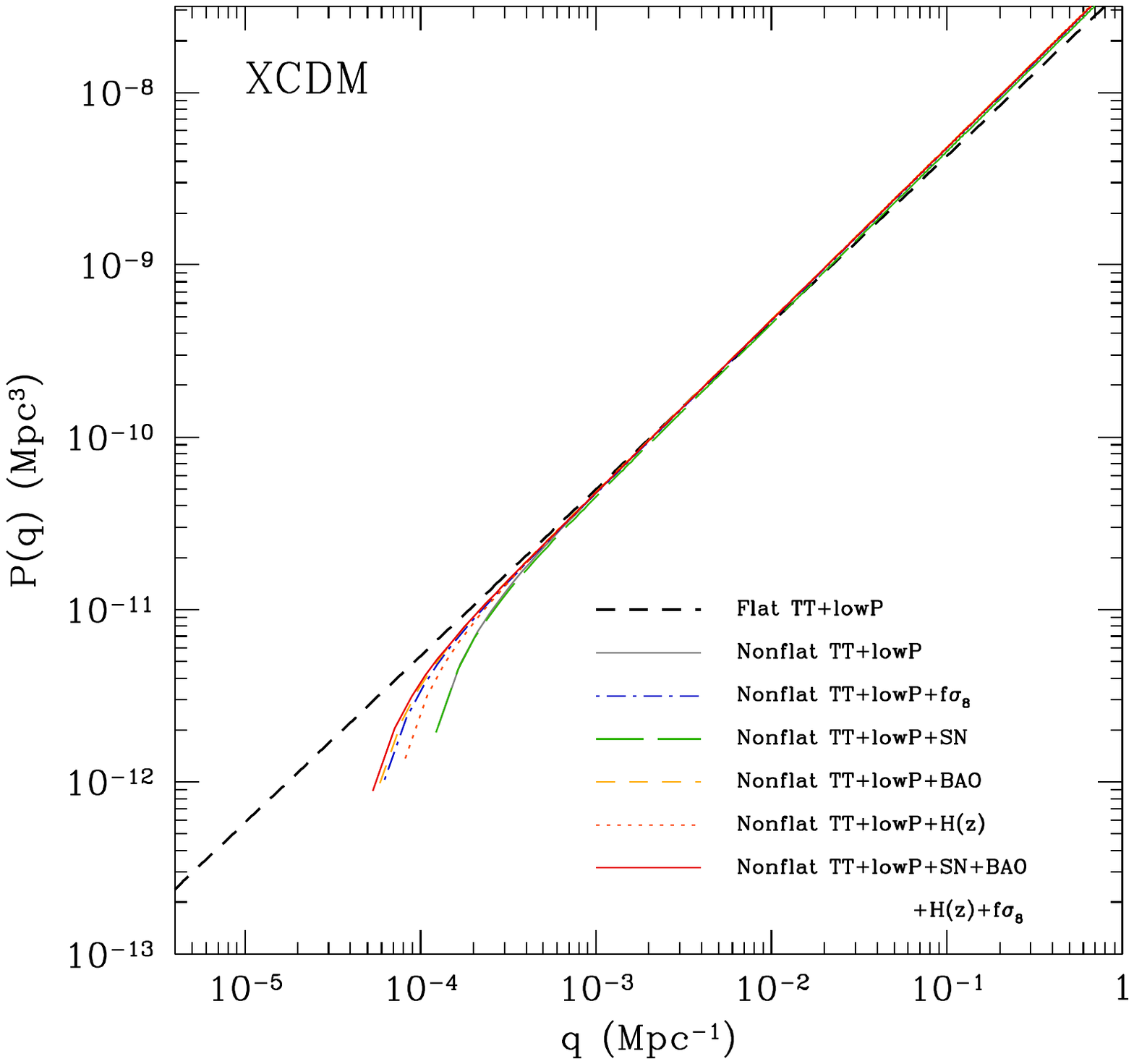}} 
\mbox{\includegraphics[width=85mm,bb=30 170 500 620]{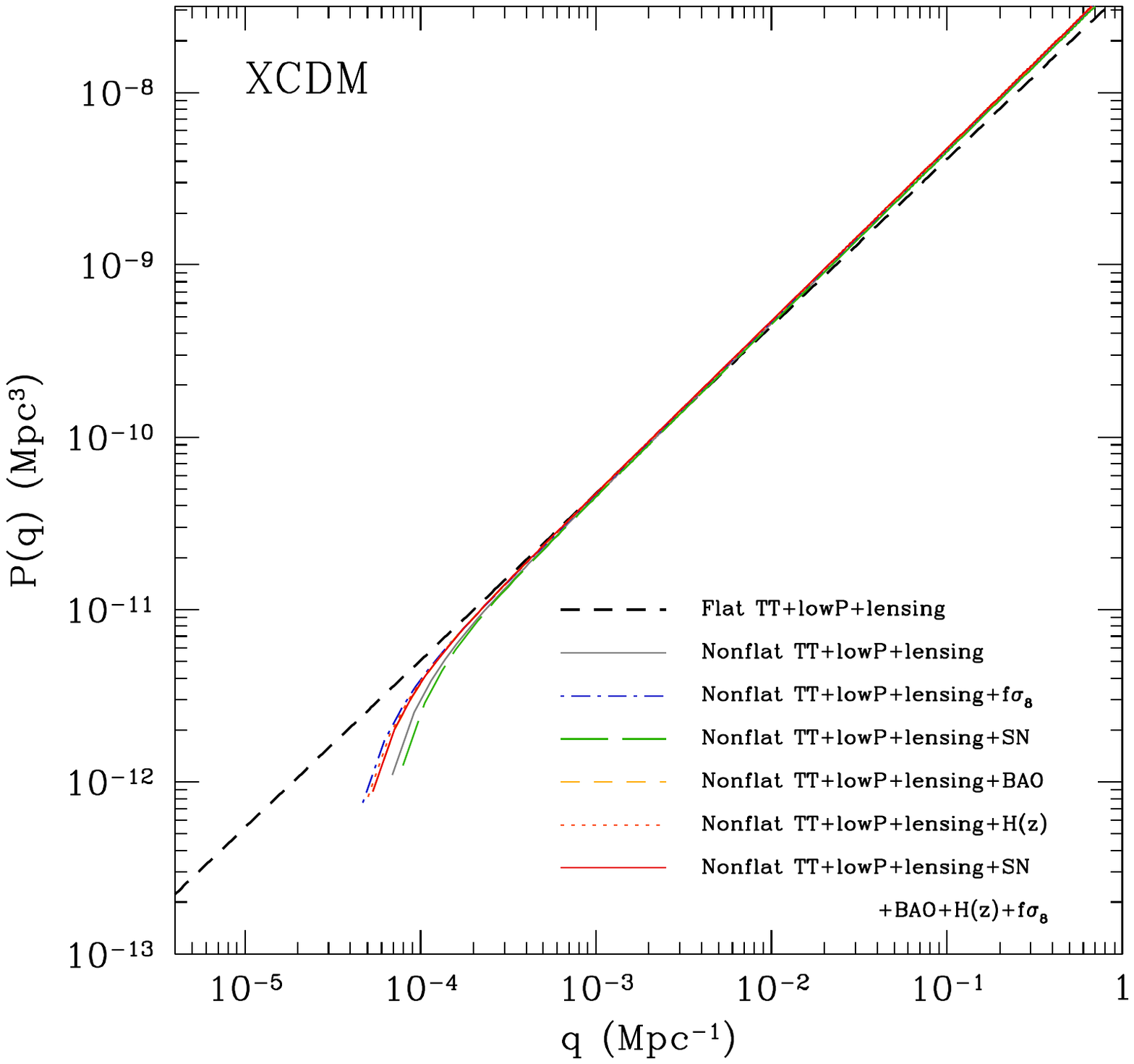}} 
\caption{Power spectra of primordial scalar-type perturbations of best-fit untilted non-power-law power spectrum nonflat XCDM cases
constrained using Planck TT + lowP data (left panel) and TT + lowP + lensing data (right panel) together
with non-CMB data sets ($f\sigma_8$, SN, BAO, $H(z)$).
In both panels the primordial power spectrum of the best-fit tilted flat-XCDM model is shown as dashed lines.
For the definition of wavenumber $q$, see Sec.\ \ref{sec:methods}.
The power spectrum is normalized to $P(q)=A_s$ at the pivot scale $k_0=0.05~\textrm{Mpc}^{-1}$. 
}
\label{fig:pq}
\end{figure*}

Figure \ref{fig:pq} shows the best-fit initial power spectra of scalar 
fractional energy density spatial inhomogeneity perturbations for the 
untilted nonflat XCDM 
parameterization constrained using the Planck TT + lowP (left) and TT + lowP + 
lensing (right panel) data in conjunction with other non-CMB data sets. The 
reduction in power at low $q$ in the best-fit closed-XCDM inflation 
parameterization spatial inhomogeneity power 
spectra shown in Fig.\ \ref{fig:pq} is partly responsible for the low-$\ell$ 
TT power reduction of the best-fit closed model $C_\ell$'s (see the 
lower panels of Figs.\ \ref{fig:ps_cmb} and \ref{fig:ps_cmb_lensing}) relative
to the best-fit tilted flat model $C_\ell$'s.\footnote{Other effects, 
including both the usual and integrated Sachs-Wolfe effects, also 
affect the shape of the low-$\ell$ $C_\ell$'s.} 
The case of the best-fit nonflat XCDM parameterization for the
TT + lowP + SN data is the most dramatic one,
consistent with the reduced low-$\ell$ TT power (Figs.\ \ref{fig:ps_cmb}$b$).

\section{Conclusion}

We measure cosmological parameters from an updated, reliable, large 
compilation of observational data by using the tilted flat-XCDM and the 
untilted nonflat XCDM dynamical dark energy inflation parameterizations. 

In summary, our main results are:
\begin{itemize}
\item We confirm, but at lower significance, the \citet{Oobaetal2018d}
result that the tilted flat-XCDM parameterization provides a better 
fit to the data than does the standard tilted flat-$\Lambda$CDM model. The
improvement is not significant, but on the other hand current data are 
unable to rule out dynamical dark energy.  
\item In the untilted nonflat XCDM case, we confirm, at higher 
significance, the \citet{Oobaetal2018b} result that cosmological data 
does not demand spatially-flat hypersurfaces for this parameterization,
and that the nonflat XCDM parameterization provides a better fit to 
the data than does the nonflat $\Lambda$CDM model (qualitatively 
it is clear that the standard tilted flat-$\Lambda$CDM model is a better 
fit to the data than is the untilted nonflat $\Lambda$CDM model).  
In the nonflat XCDM case, these data (including CMB lensing measurements) 
favor a closed model at more than 3.4$\sigma$ significance, with spatial 
curvature contributing a little less than a percent to the current 
cosmological energy budget, and favor dark energy dynamics (over a 
cosmological constant) at a little more than 1.2$\sigma$.
\item $H_0$ values measured in both models are very similar, and consistent
with many other measurements of $H_0$. However, as well known, $H_0$ 
estimated from the local expansion rate \citep{Riessetal2018} is about
3$\sigma$ larger.   
\item $\sigma_8$ values measured in both models are close to identical and 
compatible 
with the recent DES measurement \citep{DESCollaboration2018}.
\item The measured $\Omega_m$ value is more model dependent than the measured 
$\sigma_8$ value and the $\Omega_m$ value
measured using the nonflat XCDM parameterization is more consistent with 
the recent DES estimate \citep{DESCollaboration2018}. 
\item $\Omega_b h^2$, $\tau$, $\Omega_c h^2$, as well as some of the other 
measured cosmological parameter values are model dependent.
\end{itemize}

%
%
\acknowledgements{
We thank D.\ Scolnic for providing us the Pantheon data. We acknowledge 
valuable discussions with C.\ Bennett, J.\ Ooba, and D.\ Scolnic. C.-G.P.\ 
was supported by the Basic Science Research Program through the National 
Research Foundation of Korea (NRF) funded by the Ministry of Education 
(No.\ 2017R1D1A1B03028384). B.R.\ was supported in part by DOE grant 
DE-SC0019038.
}

\begin{table*}
\caption{Individual and total $\chi^2$ values for the best-fit tilted flat and untilted nonflat $\Lambda\textrm{CDM}$ inflation models.}
\begin{ruledtabular}
\begin{tabular}{lcccccccccc}
    Data sets   & $\chi_{\textrm{PlikTT}}^2$  & $\chi_{\textrm{lowTEB}}^2$  &  $\chi_{\textrm{lensing}}^2$  &  $\chi_{\textrm{SN}}^2$  & $\chi_{\textrm{BAO}}^2$  &  $\chi_{H(z)}^2$   &  $\chi_{f\sigma_8}^2$ & $\chi_{\textrm{prior}}^2$  &  Total $\chi^2$      & $\Delta\chi^2$ \\[+0mm]
 \hline \\[-2mm]
 \multicolumn{11}{c}{Tilted flat-$\Lambda\textrm{CDM}$ model} \\
  \hline \\[-2mm]
   TT+lowP                               &   763.57 & 10496.41 &      &         &       &       &       & 1.96 & 11261.93 &      \\[+1mm]
   \quad\quad +SN                        &   763.45 & 10496.50 &      & 1036.29 &       &       &       & 2.06 & 12298.31 &      \\[+1mm]
   \quad\quad +BAO                       &   764.20 & 10495.92 &      &         & 13.02 &       &       & 2.12 & 11275.25 &      \\[+1mm]
   \quad\quad +$H(z)$                    &   763.98 & 10496.36 &      &         &       & 14.89 &       & 1.70 & 11276.93 &      \\[+1mm]
   \quad\quad +$f\sigma_8$               &   766.83 & 10494.95 &      &         &       &       & 12.15 & 1.87 & 11275.80 &      \\[+1mm]
   \quad\quad +BAO+$f\sigma_8$           &   766.67 & 10494.83 &      &         & 12.64 &       & 12.40 & 1.96 & 11288.50 &      \\[+1mm]
   \quad\quad +SN+BAO                    &   764.34 & 10495.96 &      & 1036.15 & 12.93 &       &       & 2.03 & 12311.41 &      \\[+1mm]
   \quad\quad +SN+BAO+$H(z)$             &   764.33 & 10495.93 &      & 1036.15 & 12.95 & 14.83 &       & 2.03 & 12326.21 &      \\[+1mm]
   \quad\quad +SN+BAO+$H(z)$+$f\sigma_8$ &   766.68 & 10494.90 &      & 1036.02 & 12.71 & 14.79 & 12.38 & 1.88 & 12339.36 &      \\[+1mm]
 \hline\\[-2mm]
   TT+lowP+lensing                       &   766.20 & 10494.93 & 9.30 &         &       &       &       & 2.00 & 11272.44 &      \\[+1mm]
   \quad\quad +SN                        &   766.53 & 10494.83 & 9.17 & 1036.05 &       &       &       & 2.02 & 12308.59 &      \\[+1mm]
   \quad\quad +BAO                       &   766.44 & 10494.80 & 9.13 &         & 12.61 &       &       & 2.09 & 11285.07 &      \\[+1mm]
   \quad\quad +$H(z)$                    &   766.20 & 10494.92 & 9.27 &         &       & 14.83 &       & 2.04 & 11287.27 &      \\[+1mm]
   \quad\quad +$f\sigma_8$               &   768.26 & 10494.43 & 8.67 &         &       &       & 11.31 & 1.94 & 11284.62 &      \\[+1mm]
   \quad\quad +BAO+$f\sigma_8$           &   767.56 & 10494.49 & 8.71 &         & 12.59 &       & 11.80 & 2.16 & 11297.32 &      \\[+1mm]
   \quad\quad +SN+BAO                    &   766.53 & 10494.77 & 9.03 & 1036.07 & 12.61 &       &       & 2.16 & 12321.17 &      \\[+1mm]
   \quad\quad +SN+BAO+$H(z)$             &   766.51 & 10494.80 & 9.07 & 1036.07 & 12.61 & 14.81 &       & 2.14 & 12336.01 &      \\[+1mm]
   \quad\quad +SN+BAO+$H(z)$+$f\sigma_8$ &   767.61 & 10494.48 & 8.74 & 1036.01 & 12.68 & 14.79 & 11.84 & 2.04 & 12348.20 &      \\[+1mm]
   \hline \\[-2mm]
    \multicolumn{11}{c}{Untilted nonflat $\Lambda\textrm{CDM}$ model}  \\
   \hline \\[-2mm]
   TT+lowP                               &   774.34 & 10495.42 &      &         &       &       &       & 2.33 & 11272.10 & 10.17 \\[+1mm]
   \quad\quad +SN                        &   778.23 & 10497.99 &      & 1036.74 &       &       &       & 1.97 & 12314.94 & 16.63 \\[+1mm]
   \quad\quad +BAO                       &   780.27 & 10499.20 &      &         & 14.69 &       &       & 1.92 & 11296.08 & 20.83 \\[+1mm]
   \quad\quad +$H(z)$                    &   777.14 & 10500.93 &      &         &       & 17.11 &       & 1.96 & 11297.15 & 20.22 \\[+1mm]
   \quad\quad +$f\sigma_8$               &   783.38 & 10497.49 &      &         &       &       & 11.51 & 2.41 & 11294.79 & 18.99 \\[+1mm]
   \quad\quad +BAO+$f\sigma_8$           &   783.46 & 10497.40 &      &         & 14.01 &       & 10.72 & 1.81 & 11307.41 & 18.91 \\[+1mm]
   \quad\quad +SN+BAO                    &   780.65 & 10499.11 &      & 1036.11 & 14.56 &       &       & 1.86 & 12332.30 & 20.89 \\[+1mm]
   \quad\quad +SN+BAO+$H(z)$             &   782.84 & 10497.40 &      & 1036.18 & 14.06 & 16.17 &       & 1.91 & 12348.57 & 22.36 \\[+1mm]
   \quad\quad +SN+BAO+$H(z)$+$f\sigma_8$ &   781.14 & 10499.17 &      & 1036.29 & 14.17 & 16.14 & 11.32 & 1.74 & 12359.98 & 20.62 \\[+1mm]
 \hline\\[-2mm]
   TT+lowP+lensing                       &   786.87 & 10493.86 & 9.77 &         &       &       &       & 1.79 & 11292.29 & 19.85 \\[+1mm]
   \quad\quad +SN                        &   786.65 & 10494.69 & 9.19 & 1035.95 &       &       &       & 1.83 & 12328.31 & 19.72 \\[+1mm]
   \quad\quad +BAO                       &   784.19 & 10497.32 & 9.86 &         & 13.99 &       &       & 2.04 & 11307.40 & 22.33 \\[+1mm]
   \quad\quad +$H(z)$                    &   786.87 & 10496.02 & 8.66 &         &       & 16.36 &       & 2.19 & 11310.10 & 22.83 \\[+1mm]
   \quad\quad +$f\sigma_8$               &   786.41 & 10496.00 & 8.75 &         &       &       &  9.79 & 1.99 & 11302.93 & 18.31 \\[+1mm]
   \quad\quad +BAO+$f\sigma_8$           &   788.21 & 10494.90 & 8.38 &         & 13.89 &       &  9.81 & 2.11 & 11317.31 & 19.99 \\[+1mm]
   \quad\quad +SN+BAO                    &   784.76 & 10496.50 & 9.54 & 1036.23 & 13.87 &       &       & 1.89 & 12342.79 & 21.62 \\[+1mm]
   \quad\quad +SN+BAO+$H(z)$             &   784.72 & 10496.49 & 9.60 & 1036.24 & 13.84 & 16.11 &       & 1.89 & 12358.87 & 22.86 \\[+1mm]
   \quad\quad +SN+BAO+$H(z)$+$f\sigma_8$ &   786.96 & 10495.37 & 8.63 & 1036.37 & 13.81 & 16.07 &  9.77 & 1.90 & 12368.86 & 20.66 \\[+1mm]
\end{tabular}
\\[+1mm]
Note: $\Delta\chi^2$ of an untilted nonflat $\Lambda\textrm{CDM}$ model estimated for a combination of data sets represents the excess value relative to $\chi^2$
of the tilted flat-$\Lambda$CDM model for the same combination of data sets.
\end{ruledtabular}
\label{tab:chi2_lcdm}
\end{table*}

\begin{table*}
\caption{Individual and total $\chi^2$ values for the best-fit tilted flat and untilted nonflat $\textrm{XCDM}$ inflation parameterizations.}
\begin{ruledtabular}
\begin{tabular}{lcccccccccr}
    Data sets   & $\chi_{\textrm{PlikTT}}^2$  & $\chi_{\textrm{lowTEB}}^2$  &  $\chi_{\textrm{lensing}}^2$  &  $\chi_{\textrm{SN}}^2$  & $\chi_{\textrm{BAO}}^2$  &  $\chi_{H(z)}^2$   &  $\chi_{f\sigma_8}^2$ & $\chi_{\textrm{prior}}^2$  &  Total $\chi^2$      & $\Delta\chi^2$ \\[+0mm]
 \hline \\[-2mm]
 \multicolumn{11}{c}{Tilted flat-$\textrm{XCDM}$ parameterization} \\
  \hline \\[-2mm]
   TT+lowP                               &   761.85 & 10495.08 &       &         &       &       &       & 2.02 & 11258.94 & $-2.99$     \\[+1mm]
   \quad\quad +SN                        &   763.24 & 10496.38 &       & 1035.93 &       &       &       & 2.10 & 12297.64 & $-0.67$     \\[+1mm]
   \quad\quad +BAO                       &   764.30 & 10496.20 &       &         & 12.76 &       &       & 1.94 & 11275.20 & $-0.05$     \\[+1mm]
   \quad\quad +$H(z)$                    &   763.32 & 10496.10 &       &         &       & 15.00 &       & 1.76 & 11276.18 & $-0.75$     \\[+1mm]
   \quad\quad +$f\sigma_8$               &   766.16 & 10494.26 &       &         &       &       & 11.77 & 2.03 & 11274.21 & $-1.59$     \\[+1mm]
   \quad\quad +BAO+$f\sigma_8$           &   766.79 & 10495.00 &       &         & 12.11 &       & 12.20 & 2.14 & 11288.24 & $-0.26$     \\[+1mm]
   \quad\quad +SN+BAO                    &   764.46 & 10495.90 &       & 1036.02 & 13.15 &       &       & 1.88 & 12311.39 & $-0.02$     \\[+1mm]
   \quad\quad +SN+BAO+$H(z)$             &   764.33 & 10496.04 &       & 1036.09 & 13.16 & 14.82 &       & 1.80 & 12326.24 & $+0.03$     \\[+1mm]
   \quad\quad +SN+BAO+$H(z)$+$f\sigma_8$ &   766.81 & 10494.83 &       & 1036.06 & 12.60 & 14.79 & 12.12 & 2.09 & 12339.31 & $-0.05$     \\[+1mm]
 \hline\\[-2mm]
   TT+lowP+lensing                       &   766.09 & 10493.81 &  9.39 &         &       &       &       & 2.02 & 11271.31 & $-1.13$     \\[+1mm]
   \quad\quad +SN                        &   766.35 & 10494.78 &  9.24 & 1035.95 &       &       &       & 2.01 & 12308.33 & $-0.26$     \\[+1mm]
   \quad\quad +BAO                       &   767.00 & 10494.74 &  9.06 &         & 12.25 &       &       & 1.94 & 11284.99 & $-0.08$     \\[+1mm]
   \quad\quad +$H(z)$                    &   765.98 & 10494.66 &  9.37 &         &       & 14.89 &       & 2.21 & 11287.10 & $-0.17$     \\[+1mm]
   \quad\quad +$f\sigma_8$               &   767.98 & 10493.86 &  8.65 &         &       &       & 10.77 & 1.92 & 11283.18 & $-1.44$     \\[+1mm]
   \quad\quad +BAO+$f\sigma_8$           &   768.00 & 10494.68 &  8.69 &         & 11.98 &       & 11.60 & 2.01 & 11296.96 & $-0.36$     \\[+1mm]
   \quad\quad +SN+BAO                    &   766.61 & 10494.75 &  9.04 & 1036.10 & 12.58 &       &       & 2.09 & 12321.17 & $+0.00$     \\[+1mm]
   \quad\quad +SN+BAO+$H(z)$             &   766.79 & 10494.76 &  9.06 & 1036.06 & 12.65 & 14.80 &       & 1.97 & 12336.08 & $+0.07$     \\[+1mm]
   \quad\quad +SN+BAO+$H(z)$+$f\sigma_8$ &   767.76 & 10494.51 &  8.72 & 1036.06 & 12.54 & 14.80 & 11.76 & 1.97 & 12348.12 & $-0.08$     \\[+1mm]
   \hline \\[-2mm]
    \multicolumn{11}{c}{Untilted nonflat $\textrm{XCDM}$ parameterization}  \\
   \hline \\[-2mm]
   TT+lowP                               &   771.27 & 10499.28 &       &         &       &       &       & 2.55 & 11273.11 & $+1.01$     \\[+1mm]
   \quad\quad +SN                        &   773.06 & 10496.25 &       & 1036.87 &       &       &       & 1.97 & 12308.14 & $-6.80$     \\[+1mm]
   \quad\quad +BAO                       &   780.39 & 10500.33 &       &         & 12.71 &       &       & 2.10 & 11295.53 & $-0.55$     \\[+1mm]
   \quad\quad +$H(z)$                    &   774.60 & 10499.77 &       &         &       & 19.31 &       & 1.97 & 11295.64 & $-1.51$     \\[+1mm]
   \quad\quad +$f\sigma_8$               &   779.82 & 10498.34 &       &         &       &       & 11.85 & 1.75 & 11291.75 & $-3.04$     \\[+1mm]
   \quad\quad +BAO+$f\sigma_8$           &   780.68 & 10500.54 &       &         & 12.36 &       & 11.41 & 1.83 & 11306.82 & $-0.59$     \\[+1mm]
   \quad\quad +SN+BAO                    &   777.67 & 10504.10 &       & 1036.15 & 12.73 &       &       & 2.03 & 12332.68 & $+0.38$     \\[+1mm]
   \quad\quad +SN+BAO+$H(z)$             &   778.12 & 10502.60 &       & 1036.03 & 13.24 & 16.00 &       & 1.88 & 12347.88 & $-0.69$     \\[+1mm]
   \quad\quad +SN+BAO+$H(z)$+$f\sigma_8$ &   782.42 & 10499.28 &       & 1036.33 & 12.20 & 15.57 & 10.77 & 1.94 & 12358.52 & $-1.46$     \\[+1mm]
 \hline\\[-2mm]
   TT+lowP+lensing                       & 785.83   & 10495.22 &  9.99 &         &       &       &       & 2.39 & 11293.44 & $+1.15$     \\[+1mm]
   \quad\quad +SN                        & 786.69   & 10494.26 &  9.79 & 1035.91 &       &       &       & 1.97 & 12328.62 & $+0.31$     \\[+1mm]
   \quad\quad +BAO                       & 784.55   & 10498.30 &  9.15 &         & 11.66 &       &       & 1.84 & 11305.50 & $-1.90$     \\[+1mm]
   \quad\quad +$H(z)$                    & 785.33   & 10497.40 &  8.83 &         &       & 15.58 &       & 2.00 & 11309.15 & $-0.95$     \\[+1mm]
   \quad\quad +$f\sigma_8$               & 786.23   & 10497.02 &  8.69 &         &       &       &  8.40 & 2.23 & 11302.58 & $-0.35$     \\[+1mm]
   \quad\quad +BAO+$f\sigma_8$           & 785.97   & 10497.57 &  8.63 &         & 11.73 &       &  9.81 & 1.70 & 11315.41 & $-1.90$     \\[+1mm]
   \quad\quad +SN+BAO                    & 785.67   & 10497.62 &  8.87 & 1036.15 & 12.37 &       &       & 2.30 & 12342.98 & $+0.19$     \\[+1mm]
   \quad\quad +SN+BAO+$H(z)$             & 784.07   & 10498.25 &  9.33 & 1036.75 & 11.88 & 15.43 &       & 2.40 & 12358.11 & $-0.76$     \\[+1mm]
   \quad\quad +SN+BAO+$H(z)$+$f\sigma_8$ & 784.99   & 10498.46 &  8.94 & 1036.51 & 12.09 & 15.56 &  9.83 & 1.72 & 12368.10 & $-0.76$     \\[+1mm]
\end{tabular}
\\[+1mm]
Note: $\Delta\chi^2$ of tilted flat or untilted nonflat $\textrm{XCDM}$ parameterization estimated for a combination of data sets represents
the excess value relative to $\chi^2$ of the corresponding $\Lambda\textrm{CDM}$ model for the same combination of data sets.
\end{ruledtabular}
\label{tab:chi2}
\end{table*}

%
%

\def\and{{and }}
\bibliographystyle{yahapj}


\end{document}